\documentclass[10pt,letterpaper]{article}
\usepackage{opex3}
\usepackage{color}
\usepackage{mathrsfs}
\usepackage{amsmath}
\usepackage{amssymb}

\usepackage[english]{babel}
\usepackage{bm} % con questo bm=boldsymbol
\usepackage{cite}
\usepackage{color}
\usepackage{graphicx}
\usepackage{subfigure}
\usepackage{url}

\usepackage[final]{changes}
\newcommand{\R}{\mathbb{R}}

\begin{document}

%%%%%%%%%%%%%%%%%% title page information %%%%%%%%%%%%%%%%%%
\title{Modeling of strain-induced Pockels effect in Silicon}

\author{C.~L.~Manganelli,${}^{1\dag }$ P.~Pintus,${}^{2\dag}$ and C.~Bonati${}^3$}
\address {$^{\dag}$The authors contributed equally to the paper and are listed in alphabetical order\\
$^1$ \url{costanza.manganelli@sssup.it}, Scuola Superiore Sant'Anna, via G. Moruzzi 1, 56124 Pisa, Italy, \\
$^2$ \url{paolo.pintus@sssup.it}, Scuola Superiore Sant'Anna, via G. Moruzzi 1, 56124 Pisa, Italy, \\
$^3$ \url{claudio.bonati@df.unipi.it}, INFN - Sezione di Pisa, Largo B. Pontecorvo 3, 56127 Pisa, Italy}
 
%\email{${}^*$Corresponding author: c.manganelli@sssup.it}

% \homepage{http:...} %% author's URL, if desired

%%%%%%%%%%%%%%%%%%% abstract and OCIS codes %%%%%%%%%%%%%%%%

\begin{abstract*}
We propose a theoretical model to describe the strain-induced linear
electro-optic (Pockels) effect in centro-symmetric crystals. The general
formulation is presented and the specific case of the strained silicon is
investigated in detail because of its attractive properties for integrated
optics.  The outcome of this analysis is a linear relation between the second
order susceptibility tensor and the strain gradient tensor, depending
generically on fifteen coefficients. The proposed model greatly simplifies the
description of the electro-optic effect in strained silicon waveguides,
providing a powerful and effective tool for design and optimization of optical
devices.
\vspace{0.3cm}
\end{abstract*}


%%%%%%%%%%%%%%%%%%%%%%% References %%%%%%%%%%%%%%%%%%%%%%%%%
%\phantomsection
\addcontentsline{toc}{section}{References and links}
%\bibliography{bibliography}
%\bibliographystyle{ieeetr}

%%%%%%%%%%%%%%%%%%%%%%%%%%  body  %%%%%%%%%%%%%%%%%%%%%%%%%%

%%%%%%%%%%%%%%%%%%%%%%%%%%  body  %%%%%%%%%%%%%%%%%%%%%%%%%%

\section{Introduction}
In recent years, strain engineering is emerging as a new frontier in micro and
nano-technology. By varying the elastic strain it is possible to turn on
physical and chemical properties that are absent in the unstrained material. As
a result, electronic, optical, magnetic, phononic and catalytic properties of a
material can be tuned by compressive or tensile
stress~\cite{LI2014,YILDIZ2014}.

In optics, tensile strained germanium and strained silicon are attracting a
great deal of interest. Tensile strained germanium-on-silicon can be used as
active material for the short-wavelength infrared light and it can be an
efficient solution for manufacturing monolithic lasers and optical
amplifiers~\cite{LIU2010, VIRGILIO2013}. On the other hand, Pockels effect has
been experimentally  measured in strained silicon ~\cite{JACOBSEN2006}, making
it a promising candidate material for realizing very fast integrated optical
modulators and switches. The modeling of strain-induced electro-optic Pockels
effect in silicon is the main object of this work.

The electro-optic effect consists in the change of the refractive index induced
by an electric field that varies slowly compared with the frequency of an
optical signal~\cite{YARIV1984_OE, LIU2009}. In the particular case of the
Pockels effect, also called linear electro-optic effect, the change in the
refractive index is proportional to the applied electric field, providing an
efficient physical mechanism for optical modulation. However, a peculiarity of
the Pockels effect is that it arises only in crystalline solids lacking of
inversion symmetry~\cite{YARIV1984}. As a consequence, for centro-symmetric
crystals (like silicon) the Pockels effect can be observed only when the
inversion symmetry is broken, e.g., by the presence of significant
surface/interface effects~\cite{HUANG1994, MITCHELL2001,ZHAO2009} or by an
inhomogeneous mechanical stress.

Since the seminal work~\cite{JACOBSEN2006} there have been considerable
progresses in the fabrication of electro-optic modulators based on strained
silicon. In 2011 the first fully integrated Mach-Zehnder interferometer (MZI),
based on  strained silicon rib waveguides, was manufactured and the value of
122\,pm/V for the effective electro-optic susceptibility was  measured
~\cite{CHMIELAK2011}. Two years later, the same authors presented a detailed
investigation of the local strain distribution and of the induced optical
nonlinearity as a function of the waveguide width, measuring the record value
of 190\,pm/V for a 300\,nm large rib waveguide~\cite{CHMIELAK2013}.
The dependence of the second order dielectric
susceptibility on both wavelength and waveguide width was later investigated
for a channel waveguide, showing that higher values of the effective susceptibility can
be reached for narrow waveguides and large wavelength ~\cite{DAMAS2014}. 

The more recent analysis performed in \cite{SHARIF2015}
and \cite{SHARMA2015_arxiv}suggests that the phase
shift observed in previous MZI experiments (and thus the corresponding effective index
variation) can be also related to the presence of free carrier variation inside the
waveguide. The inversion of the phase shift observed when switching the applied
tension, which was previously interpreted as the smoking-gun signal for the
Pockels effects, is now attributed also to the surface charge present in the
silicon nitride cladding.  The results presented in \cite{SHARMA2015}, where
the free-carrier plasma dispersion effect in silicon waveguides has been
theoretically characterized, further support the possibility that the results
obtained in \cite{CHMIELAK2011, CHMIELAK2013} and \cite{DAMAS2014} suffer from
contamination of free-carrier contribution.

The most natural way to avoid contamination from free-carriers would be to
perform high-speed measurements, with temporal resolution smaller than the
free-carrier response times.  In particular, an unambiguous indication for the
presence of a strain-induced second order susceptibility is given by second
harmonic generation (SHG) measurements.  The experimental value of the second
order dielectric susceptibility extracted from SHG measurements is of 40\,pm/V
\cite{CAZZANELLI2011}, which is expected to be of the same order of magnitude
of the one related to the Pockels effects. The comparison of these two effects
can however be only qualitative, since they are associated with different
frequency components of the nonlinear susceptibility.

On the one hand, when the free-carrier concentration are known, 
the corresponding effective index variation can be predicted by 
the empirical formula proposed by Soref and Bennett in 1987~\cite{SOREF1987}, and 
latterly improved by Nedeljkovic et al.~\cite{NEDELJKOVIC2011}. On the other hand
several approaches have been proposed 
to model the strain-induced Pockels effect in silicon, however 
no one can be effectively used for practical purpose.
   
In~\cite{HON2009} it is shown that a simplified classical model of a 2D
centro-symmetric lattice is not able to reproduce the correct order of magnitude
of the experimentally measured susceptibilities, moreover, in this approach,
the way in which the strain enters the computations does not appear to be
completely justified. A more sophisticated model is used in
\cite{CAZZANELLI2011}, where the linear electro-optic effect in strained
silicon is studied by using the time-dependent density-functional
theory~\cite{LUPPI2010, LUPPI2010_PRB}.  While this {\it ab initio} method is
theoretically well founded, it has the obvious drawback of being
computationally very expensive.

Some intuitive ideas already present in the literature relate the
effective susceptibility to the strain gradient (see, e.g.,
\cite{CAZZANELLI2011, CHMIELAK2013, DAMAS2014, PUCKETT2014}). Indeed, as
explicitly noted in \cite{DAMAS2014}: ``Despite the lack of general proof for
this claim available in the literature yet, it has been widely accepted that
the second order nonlinear effects in strained silicon are caused by the
variations of strain i.e. strain gradients inside the crystal.'' In this work
we will show that such a relation can be deduced by using just symmetry
arguments and the specific case of the strained silicon will be investigated as
a particularly interesting example. The final result will be a simple linear
relation between the second order effective susceptibility tensor and the
strain gradient tensor (weighted by the electromagnetic modes), depending 
generically on fifteen independent coefficients that can in principle be
obtained from experimental measurements.
Once these coefficients are known, the computation of the electro-optic effect
is reduced to a standard strain computation and electromagnetic mode analysis,
thus providing an easy framework for the optimization of optical devices.

\section{Nonlinear susceptibility and Pockels effect}\label{sec:electro-optic}

In this section we recall, for the benefit of the reader and to fix the
notation, the basic properties of the quadratic nonlinear susceptibility that
will be needed in the following. 

The quadratic nonlinear susceptibility is conventionally defined, for a local
and causal medium, by the following relation between the (quadratic component
of the) polarization vector $\bm{P}^{(2)}(t)$ and the electric field
$\bm{E}(t)$: \begin{equation}\label{Eq:1}
P_i^{(2)}(t)=\epsilon_0\iint\limits_0^{\quad\ \infty}\chi_{ijk}^{(2)}(\tau_1,\tau_2)
E_j(t-\tau_1)E_k(t-\tau_2)\mathrm{d}\tau_1\mathrm{d}\tau_2\ ,
\end{equation}
where $\epsilon_0$ is the free-space dielectric permittivity,
$\chi^{(2)}_{ijk}$ is the susceptibility tensor. Sum over repeated indices is
always understood  in Eq.~\eqref{Eq:1} and
followings unless otherwise explicitly stated. The dependence on the position $\bm{x}$ is omitted for the sake of
the simplicity but will be important in the following sections. The previous
relation can be rewritten in the frequency domain as following
\begin{equation}
P_i^{(2)}(\omega_1+\omega_2)=\epsilon_0\chi_{ijk}^{(2)}(\omega_1+\omega_2;\omega_1,\omega_2)
E_j(\omega_1)E_k(\omega_2)\ ,
\end{equation}
where the first argument of the susceptibility is just the sum of the two other
frequencies, a notation conventionally adopted in the literature. 

The symmetry properties of the tensor
$\chi_{ijk}^{(2)}(\omega_1+\omega_2;\omega_1,\omega_2)$ will be
particularly important in our analysis. From the definition and the reality
of the fields it easily follows that
\begin{equation}
\chi_{ijk}^{(2)}(\omega_1+\omega_2;\omega_1,\omega_2)=
\chi_{ikj}^{(2)}(\omega_1+\omega_2;\omega_2,\omega_1)
\end{equation}
and
\begin{equation}
\chi_{ijk}^{(2)}(\omega_1+\omega_2;\omega_1,\omega_2)=
\chi_{ijk}^{(2)}(-\omega_1-\omega_2;-\omega_1,-\omega_2)^*.
\end{equation}
For a lossless and weakly dispersive medium, when no external static magnetic
fields are present, it can be shown  that
$\chi_{ijk}^{(2)}(\omega_1+\omega_2;\omega_1,\omega_2)$ is real and it is
invariant under any permutation of the indices, provided the arguments are
similarly permuted (see e.g. \cite{BOYD2010}  for an explicit computation in
perturbation theory or~\cite{Landau8} for an indirect argument). In particular,
in the limit of zero frequencies it is symmetric under a generic permutation of
the indices. 

The Pockels effect consists in the variation of the index of refraction at
frequency $\omega$ when a static electric field is applied. To investigate the
Pockels effect we thus have to study $\bm{P}^{(2)}(\omega)$ when the electric
field is $\bm{E}(t)=\bm{E}^{dc}+\R e[\bm{E}^{opt}(\omega)e^{-i\omega t}]$,
where we denoted by $\bm{E}^{dc}$ the static (real) electric field and by
$\bm{E}^{opt}(\omega)$ the component of frequency $\omega$ of the optical
field. The relevant frequency components of the quadratic susceptibility are
thus $\chi_{ijk}^{(2)}(\omega; \omega, 0)$ and the general symmetries previously
discussed become now
\begin{equation}\label{eq:chi_pockels}
\begin{aligned}
& \chi_{ijk}^{(2)}(\omega; \omega, 0) =\chi_{ikj}^{(2)}(\omega; 0, \omega)\ , \\
& \chi_{ijk}^{(2)}(\omega; \omega, 0) =\chi_{ijk}^{(2)}(-\omega; -\omega, 0)
=\chi_{jik}^{(2)}(\omega; \omega, 0)\ ,
\end{aligned}
\end{equation}
thus $\chi_{ijk}^{(2)}(\omega; \omega, 0)$ is invariant under permutation of
the first two indices. Using these properties it is simple to obtain the
following form for the polarization in the case of the Pockels effect
\begin{equation}\label{eq:pockels_p}
P_i^{(2)}(\omega)= 2\chi^{(2)}_{ijk}(\omega;\omega, 0) E_j^{opt}(\omega)E_k^{dc}\ .
\end{equation}
For  the sake of simplicity, we will suppress in the following the superscript of
the optical field $\bm{E}^{opt}$ that will be simply denoted by $\bm{E}$.   
The phenomenon of second harmonic generation can also be described using
similar relations, the main difference being that the relevant
frequency components of the second order susceptibility are in that case
$\chi_{ijk}^{(2)}(2\omega; \omega,\omega)$, wich are symmetric under the
exchange of the last two indices.

Note that, in addition to the previously discussed symmetries of the quadratic
susceptibility, the lattice symmetry must also be added in the case of
crystals: the tensor $\bm{\chi}^{(2)}$ is invariant under the symmetry group of
the lattice. In particular, for centro-symmetric lattices (that are invariant
under the inversion symmetry $\bm{x}\to -\bm{x}$) the tensor $\bm{\chi}^{(2)}$
has to vanish, like all the invariant tensors with an odd number of indices.

In the following, when discussing lattice symmetries, it will obviously be
convenient to work in the crystallographic frame, however attention has to be
paid to the fact that this frame typically does not coincide with the device
coordinate one, so that a change of reference frame is needed to obtain
expressions of direct physical application~\cite{ESSENI2011}.

\section{The strain-induced Pockels effect}\label{sec:model}

In the linear theory of elasticity, a small deformation
$\bm{x}\to\bm{x}+\bm{u}(\bm{x})$ is described by the symmetric strain tensor
$\bm{\varepsilon}$, defined by
\begin{equation}
\varepsilon_{ij}=\frac{\partial u_i}{\partial x_j}+\frac{\partial u_j}{\partial x_i}\ 
\end{equation}
where $\bm{u}(\bm{x})$ represents the displacement of a material
point~\cite{LANDAU7}. In order to determine the relation between
$\bm{\chi}^{(2)}$ and $\bm{\varepsilon}$ we will follow the same philosophy of
effective field theories in theoretical physics: the most general expression
compatible with the symmetries of the problem is considered, then the various
possible terms are classified according to their ``strength'' retaining only
the most relevant ones. The final relation will depend on a number of unknown
constants, which are usually called ``low energy constants'' and have to be
fixed by comparing with experimental data.

As a starting point we assume that $\bm{\chi}^{(2)}$ is a local functional of
$\bm{\varepsilon}$, i.e. that $\bm{\chi}^{(2)}$ at point $\bm{x}$ depends only
on the values of $\bm{\varepsilon}$ and its derivatives at $\bm{x}$ (the
possible dependence of observable quantities on the strain gradient tensor is
well known in the litature, see e.g.  \cite{FLECK1997}). We then assume that
this dependence is analytic and we develop everything in Taylor series, thus
arriving to the following expression \begin{equation}\label{eq:chi_general}
\begin{aligned}
\chi_{ijk}^{(2)}&=\chi_{ijk}^{(2)}|_{\varepsilon=0}
+\left.\frac{\partial \chi_{ijk}^{(2)}}{\partial \varepsilon_{\alpha\beta}}\right|_{\varepsilon=0}
\hspace{-0.5cm}\varepsilon_{\alpha\beta}
+\left.\frac{\partial \chi_{ijk}^{(2)}}{\partial\zeta_{\alpha\beta\gamma}}\right|_{\varepsilon=0}
\hspace{-0.5cm}\zeta_{\alpha\beta\gamma} %\\&
+\left.\frac{\partial^2 \chi_{ijk}^{(2)}}{\partial \varepsilon_{\alpha\beta}
\partial\varepsilon_{\gamma\delta}}\right|_{\varepsilon=0}
\hspace{-0.5cm}\varepsilon_{\alpha\beta}\varepsilon_{\gamma\delta}\\&
+\left.\frac{\partial^2 \chi_{ijk}^{(2)}}{\partial \varepsilon_{\alpha\beta}
\partial\zeta_{\gamma\delta\mu}}\right|_{\varepsilon=0}
\hspace{-0.5cm}\varepsilon_{\alpha\beta}\zeta_{\gamma\delta\mu}%\\&
+\left.\frac{\partial^2 \chi_{ijk}^{(2)}}{\partial\zeta_{\alpha\beta\gamma}
\partial\zeta_{\delta\mu\nu}}\right|_{\varepsilon=0}
\hspace{-0.5cm}\zeta_{\alpha\beta\gamma}\zeta_{\delta\mu\nu}
+\cdots\ ,
\end{aligned}
\end{equation}
where we introduced the shorthand 
\begin{equation}
\zeta_{\alpha\beta\gamma}=\frac{\partial \varepsilon_{\alpha\beta}}{\partial x_{\gamma}}
\end{equation}
and where dots stand for terms involving higher derivatives of the strain
tensor and higher orders of the Taylor expansion.  The subscript
$|_{\varepsilon=0}$ means that the derivatives have to be computed at vanishing
deformation.  

If we now specialize to the case of centro-symmetric crystals, all the
coefficients with an odd number of indices identically vanish, thus the
previous expression becomes
\begin{equation}\label{eq:chi_general2}
\begin{aligned}
\chi_{ijk}^{(2)}&=\left.\frac{\partial \chi_{ijk}^{(2)}}{\partial\zeta_{\alpha\beta\gamma}}\right|_{\varepsilon=0}
\hspace{-0.5cm}\zeta_{\alpha\beta\gamma} +\left.\frac{\partial^2 \chi_{ijk}^{(2)}}{\partial \varepsilon_{\alpha\beta}
\partial\zeta_{\gamma\delta\mu}}\right|_{\varepsilon=0}
\hspace{-0.5cm}\varepsilon_{\alpha\beta}\zeta_{\gamma\delta\mu}
+\cdots\ .
\end{aligned}
\end{equation}
In particular all the terms depending on $\bm{\varepsilon}$ but not on its
derivatives disappear.  This is a direct consequence of the fact that a uniform
strain does not break the inversion symmetry and thus cannot induce a
non-vanishing quadratic susceptibility. 

The various term in the right hand side of Eq.~\eqref{eq:chi_general2} can be
classified according to their power of the strain and their number of
derivatives. In the limit of small deformation and  (by writing explicitly the
dependence on the frequencies and on the position) the leading contribution is 
\begin{equation}\label{eq:T_definition}
\chi_{ijk}^{(2)}(\bm{x}; \omega_1+\omega_2;\omega_1,\omega_2)=
T_{ijk\alpha\beta\gamma}(\omega_1+\omega_2;\omega_1,\omega_2)
\zeta_{\alpha\beta\gamma}(\bm{x})
\end{equation}
and we thus expect a linear relation between the tensors $\bm{\chi}^{(2)}$ and $\bm{\zeta}$.
 
The tensor $\bm{T}$ inherits some symmetries from $\bm{\chi}^{(2)}$ and
$\bm{\varepsilon}$: it is symmetric for $\alpha\leftrightarrow\beta$ and, as
far as the Pockels effect is concerned, for $i\leftrightarrow j$ (see
Eqs.~\eqref{eq:chi_pockels}). It is thus useful to adopt the contracted index
notation $\hat{T}_{\{ij\}k\{\alpha\beta\}\gamma}$ (see
App.~\ref{sec:compacted}), that points out the fact that only $324$ of the
$3^6=729$ components of $\bm{T}$ are linearly independent. Since $\bm{T}$ is an
invariant tensor for the lattice symmetry, this number can be further largely
reduced. In the case of the lattice octahedral symmetry typical of the silicon
crystal \cite{YARIV1984}, the number of independent components is in fact only
$15$. The general procedure to identify the independent elements is reported in
App.~\ref{sec:TSymmetriesAlg}, while  the explicit form of
Eq.~\eqref{eq:T_definition} in the crystal reference frame is given in
App.~\ref{sec:explicit_form}.

\section{The effective susceptibility}\label{sec:chi_eff}

We have shown in the previous section that a linear relation between the
$\bm{\chi}^{(2)}(\bm{x})$ tensor and the strain gradient $\bm{\zeta}(\bm{x})$
has to be expected, however this local relation is not easily accessible by
experiments. What is typically measured (see
\cite{CHMIELAK2011, CHMIELAK2013, DAMAS2014}) is the variation of the
effective refractive index $n^{\mathrm{eff}}$ for a propagating waveguide mode
induced by the switching-on of the static  electric field. The change of the
effective refractive index due to the Pockels effect can easily be obtained
(see App.~D) and it is given by
\begin{equation}\label{eq:delta_neff}
\Delta n^{\mathrm{eff}}=\frac{\epsilon_0 c}{N}\int_{A}E_i^{\,*}
\chi^{(2)}_{ijk}(\omega;\omega,0)E_jE^{dc}_k\,\mathrm{d} A\ ,
\end{equation}
where the spatial dependence of the electric fields and of $\bm{\chi}^{(2)}$ is
implied, $A$ is the cross-section of the silicon waveguide and the
normalization factor $N$ is the active power of the optical mode that
propagates along the waveguide, given by 
\begin{equation}\label{eq:active_power}
N=\frac{1}{2}\int_{A_{\infty}}(\bm{E}\times\bm{H}^*+\bm{E}^{\,*}\times\bm{H})\cdot\bm{i}_z\,\mathrm{d}A\ ,
\end{equation}
where $\bm{i}_z$ denotes the unit vector parallel to the
direction of propagation and $A_{\infty}$ is the whole plane orthogonal to the
waveguide.  In a typical experimental setup, the metal contacts are much larger
that the optical waveguide and the electrostatic field can be assumed constant
in Eq.~\eqref{eq:delta_neff}~\cite{CHMIELAK2011,CHMIELAK2013,DAMAS2014}.

In order to compare different experimental results, obtained using different
devices, it is convenient to introduce an effective susceptibility that relates
$\Delta n^{\mathrm{eff}}$ to $\bm{E}^{dc}$.  It is worth noting that, since
$\Delta n^{\mathrm{eff}}$ is a scalar quantity, an effective susceptibility
defined in such a way will not be a $3-$index tensor like $\bm{\chi}^{(2)}$,
but simply a vector. Another aspect that is usually not fully appreciated is
the fact that the effective susceptibility  not only provides a quantitative
estimate of the non-vanishing properties of $\bm{\chi}^{(2)}$, but it is also
related to the effectiveness of the device in maximizing $\Delta
n^{\mathrm{eff}}$. 

There are several definition of the effective susceptibility that are used in
the literature and indeed, for this reason, some care is required in comparing
results from different experimental groups (like, e.g.,
\cite{CHMIELAK2011, CHMIELAK2013} and \cite{DAMAS2014}). The simplest
definition  is
just given by
\begin{equation}\label{eq:chi_eff}
\chi^{\mathrm{eff}}_kE^{dc}_k=n^{\mathrm{eff}}\Delta n^{\mathrm{eff}}\ ,
\end{equation}
where the index $k$ is the direction of the static field, like in
Eq.~\eqref{eq:delta_neff}. When only one component of the static field is
different from zero, it is straightforward to derive the corresponding entry
of the effective susceptibility vector. 
The definition in Eq. \eqref{eq:chi_eff} appears to be the natural generalization
to anisotropic optical waveguides of the simple result $n\Delta n=
\chi^{(2)}E^{dc}$, which is valid for the Pockels effect in homogeneous and
isotropic materials, and it follows from $n^2=\epsilon/\epsilon_0$ and
$\epsilon=\epsilon_0(1+\chi^{(1)}+2\chi^{(2)}E^{dc})$. 

Using in this definition of effective susceptibility
the relation of Eq.~\eqref{eq:T_definition} leads to
\begin{equation}\label{eq:chi_eff_zeta}
\chi^{\mathrm{eff}}_k(\omega;\omega,0)=T_{ijk\alpha\beta\gamma}(\omega;\omega,0)\, 
\overline{\zeta_{\alpha\beta\gamma}^{ij}}(\omega)\ ,
\end{equation}
where we defined the weighted strain gradient as
\begin{equation}\label{eq:weighted_zeta}
\overline{\zeta_{\alpha\beta\gamma}^{ij}}(\omega)=\frac{\epsilon_0 c\,
n^{\mathrm{eff}}}{N}\int_{A}E_i^{\,*}(\bm{x})
\zeta_{\alpha\beta\gamma}(\bm{x})E_j(\bm{x})\,\mathrm{d} A\ .  
\end{equation}
It can be noted that the frequency dependence of
$\chi^{\mathrm{eff}}_k(\omega;\omega,0)$ originates from several sources: the
possible explicit dependence of $\bm{T}(\omega;\omega,0)$ on $\omega$, the
presence of $n^{\mathrm{eff}}$ in the definition of the weighted strain
gradients and, implicitly, through the form of the optical waveguide mode
$\bm{E}(\bm{x})$ in Eq.~\eqref{eq:weighted_zeta}.

Since the experimentally observed dependence of $\chi^{\mathrm{eff}}$ on the
frequency is very mild (see \cite{DAMAS2014}), it is reasonable to
 assume that the tensor $\bm{T}$ is
frequency independent, i.e. that all the dependence of $\chi^{\mathrm{eff}}_k$
on $\omega$ can be explained by the dependence of the weighted strain gradient
on $\omega$.  It should be clear that this is not expected to be always true
and this assumption can only be justified/disproved by its ability/inability to
reproduce experimental data.  When new, more precise, data will appear, it is
conceivable that this hypothesis will have to be relaxed.

\section{Investigation of strain-induced susceptibility in fabricated devices}
In this section we investigate two waveguide fabricated and characterized in \cite{CHMIELAK2011,CHMIELAK2013} and \cite{DAMAS2014}, respectively. The 15 independent entries of tensor \bm{$T$} can be potentially computed by fitting the experimental results. However, because the results in the literature might be contaminated by the free carrier effect \cite{SHARIF2015,SHARMA2015}, we limit our analysis to some general properties of the strain-induced $\bm{\chi^{(2)}}$, showing the role played by the geometry to provide different effective second order susceptibility.

\subsection{Devices under investigation}
\begin{figure}[!ht]
\centering{
\subfigure[
\label{fig:Chmielak}]{\includegraphics[width=0.45\columnwidth]{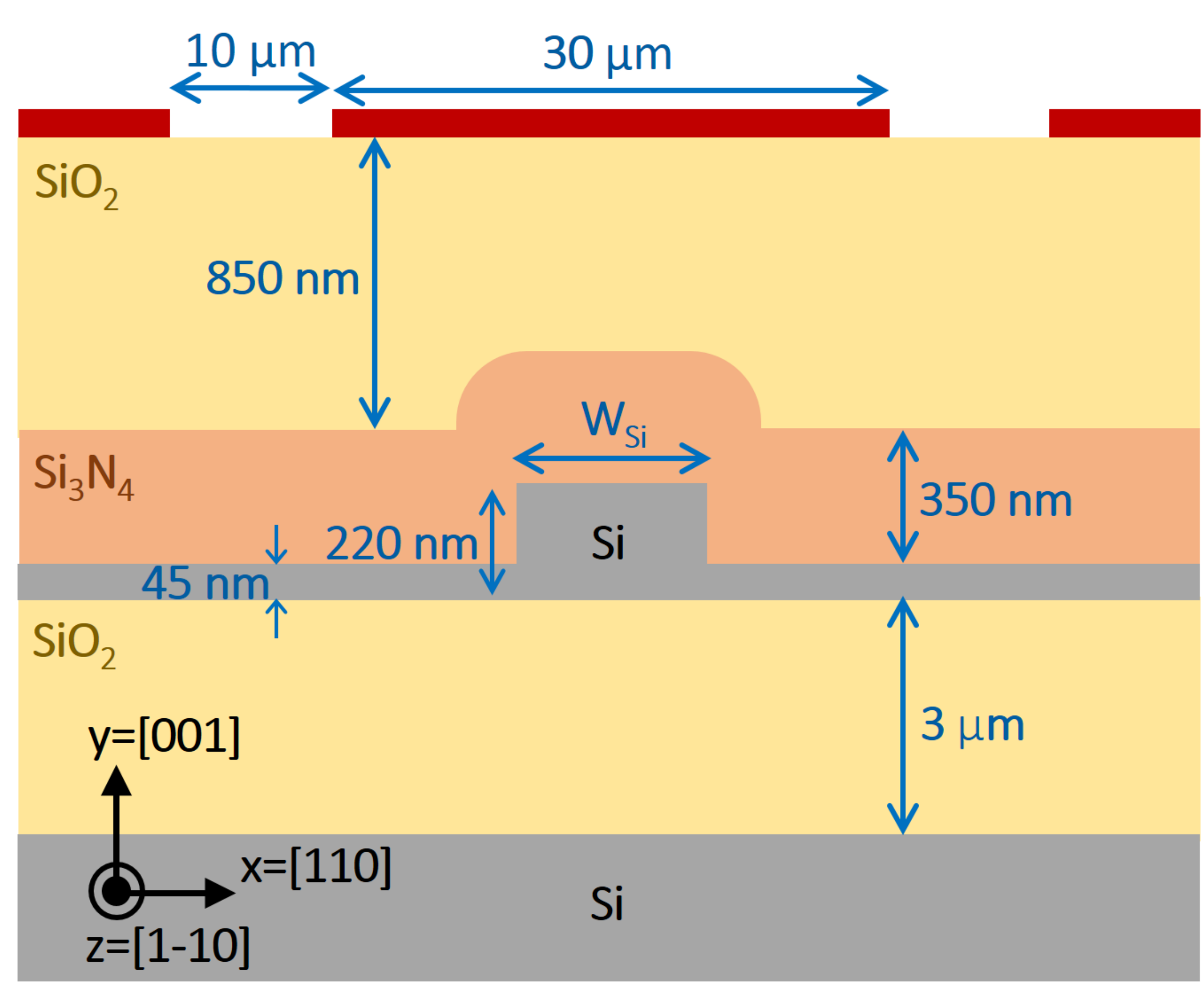}}
\quad
\subfigure[
\label{fig:Damas}]{\includegraphics[width=0.45\columnwidth]{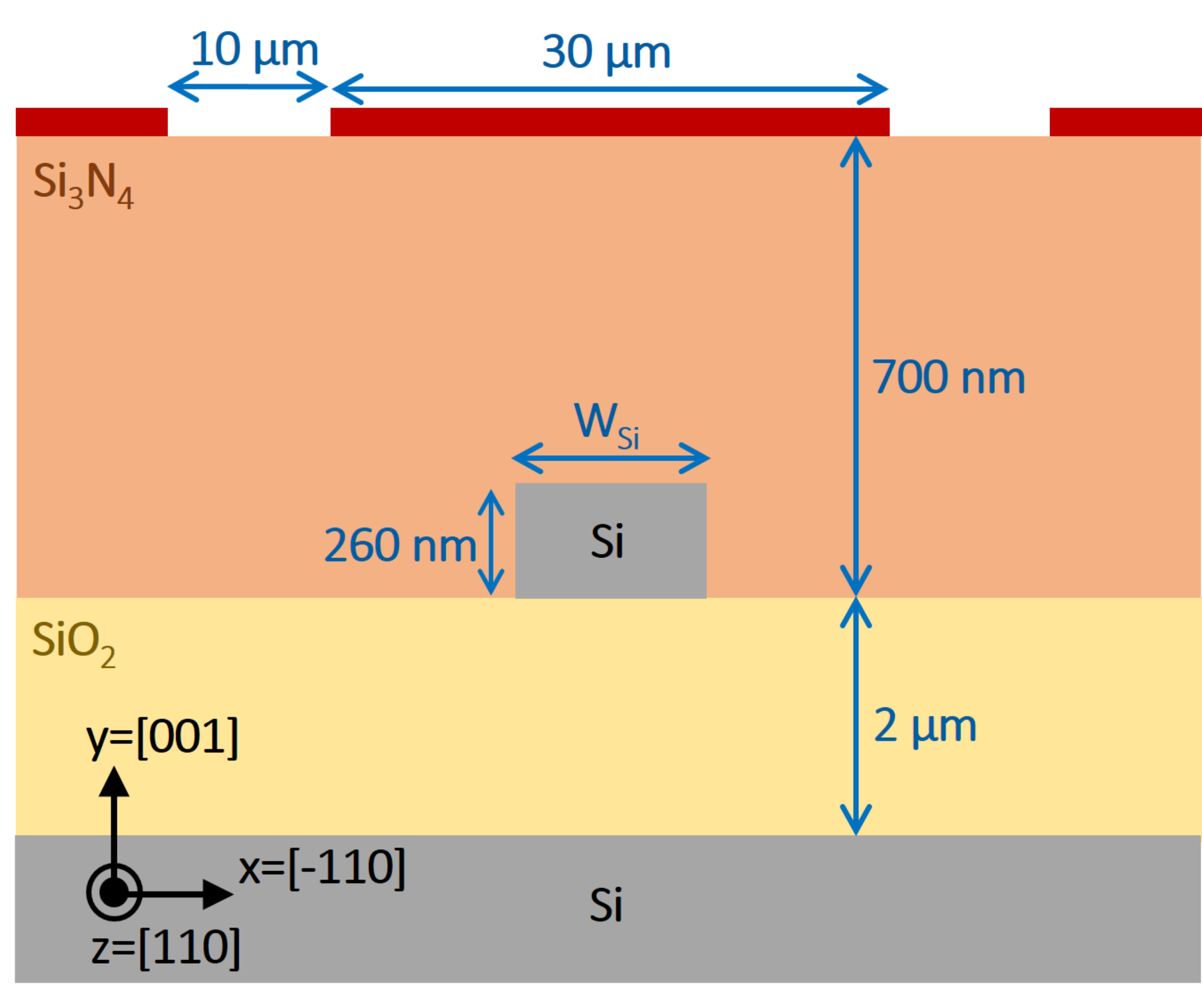}}
\caption{Cross-section of the strained silicon based MZI studied
in~\cite{CHMIELAK2011,CHMIELAK2013} and \cite{DAMAS2014}, respectively. 
 In (a) the slab waveguide cross-section described in
\cite{CHMIELAK2011,CHMIELAK2013}.  The cases of waveguide width ($w_{Si}$) equal
to 300\,nm, 350\,nm, 400\,nm, 450\,nm, and 500\,nm have been investigated.
 In (b) the channel waveguide cross-section described
in~\cite{DAMAS2014}. The cases of waveguide width ($w_{Si}$) equal to 385\,nm,
435\,nm, and 468\,nm have been investigated.
Pictures are not to scale.}\label{fig:MZI_Fig}}
\end{figure}

The waveguide cross-section used in~\cite{CHMIELAK2011,CHMIELAK2013} is
schematically shown in Fig.~\ref{fig:Chmielak}. A silicon rib waveguide was
manufactured on a silicon-on-insulator (SOI) substrate with a 220\,nm thick
(100)-oriented top silicon layer over a 3\,$\mu$m thick buried oxide. Silicon
waveguides are fabricated by an etching process that leaves a slab thickness of
45\,nm. A 350\,nm thick Si$_3$N$_4$ layer is deposited using remote plasma
enhanced chemical vapor deposition and after the annealing process, a
protective $\mathrm{SiO}_2$ cladding layer is deposited on the top (850\,nm
thick).

The geometry adopted in~\cite{DAMAS2014} is slightly different and it is
schematically shown in Fig.~\ref{fig:Damas}. A fully etched channel waveguide
is fabricated on a 260\,nm thick (100)-oriented top silicon layer with
2\,$\mu$m buried oxide. The silicon waveguide is eventually covered by a single
700\,nm layer of $\mathrm{Si}_3\mathrm{N}_4$. This choice was motivated by the
fact that the protective $\mathrm{SiO}_2$ cladding layer on
$\mathrm{Si}_3\mathrm{N}_4$ was observed to reduce the overall stress in the
silicon wafer and the induced  \hyphenation{nonlinearity}  nonlinearity.

\subsection{Strain simulation details}

In order to investigate the consequences of Eqs.~\eqref{eq:chi_eff_zeta} and
\eqref{eq:weighted_zeta}, we need to evaluate the strain gradient for the
structures described in the previous section. Since silicon elastic properties
significantly depend on the orientation of the crystalline structure, they have
been taken into account in the mechanical simulation.
To describe the deformation of silicon, one possibility could be to make use of
the Hooke's law, i.e. the linear relation between strain and stress, which for
materials with cubic symmetry involves only three independent
components~\cite{HOPCROFT2010}. A more convenient description, that avoids
tensorial transformation, is however the one that makes use of the orthotropic
model~\cite{HOPCROFT2010}. A material is said to be orthotropic when it has at least two
orthogonal planes of symmetry. Its elasticity can be described by a matrix that
takes into account the fundamental elasticity quantities in the axes of
interest: the Young's module ($Y$), the Poisson's ratio ($\nu$) and the shear
modulus ($G$). In this work we use the letter $Y$ for the Young's module
instead of $E$, traditionally used, to avoid confusion with the electric field.

The most common use of orthotropic expressions for silicon is to provide the
elasticity values in the frame of a standard (100)-silicon wafer. When
$z=[110]$, $x=[\bar{1}10]$, $y=[001]$, like in the device investigated in
\cite{DAMAS2014} the elasticity moduli are~\cite{HOPCROFT2010}
\begin{equation}\label{eq:elastic-100wafer}
\begin{aligned}
& Y_x=169\,\mathrm{GPa}    & & Y_y=130\,\mathrm{GPa}       & & Y_z=Y_x  \\
& \nu_{xy}=0.36            & & \nu_{yz}=0.28 		 & & \nu_{xz}=0.064\\
& G_{xy}=79.6\,\mathrm{GPa}& & G_{yz}=G_{xy} & & G_{xz}=50.9\,\mathrm{GPa} .
\end{aligned}
\end{equation}
Similar relation can be derived for the device in
\cite{CHMIELAK2011,CHMIELAK2013}, where $x$ and $z$ are switched in
Eq.~\eqref{eq:elastic-100wafer}.

The deformation of the silicon waveguides has been computed using COMSOL
multi-physics tools~\cite{COMSOL}. Assuming 1\,GPa compressive stress as the
initial condition for the silicon nitride layer, the elastic strain has been
computed for the structures shown in Fig.\ref{fig:Chmielak} and
Fig.\ref{fig:Damas} as a function of the waveguide width and the wavelength.  For the silicon \hyphenation{waveguide} waveguide and
the buried silicon we used the values of the elastic modulus, the shear
modulus and the Young's modulus in Eqs.~\eqref{eq:elastic-100wafer}. Since for
the solid analysis a 2D model has been considered, the components
$\varepsilon_{xz}$ and $\varepsilon_{yz}$ of the strain identically vanish; for
the same reason, the derivative of the strain coefficients with respect to $z$
are assumed equal to zero. 

\begin{figure}[!ht]
\centering
\subfigure[\protect \label{fig:exx_senzacrac}]{\includegraphics[width=0.4\columnwidth]{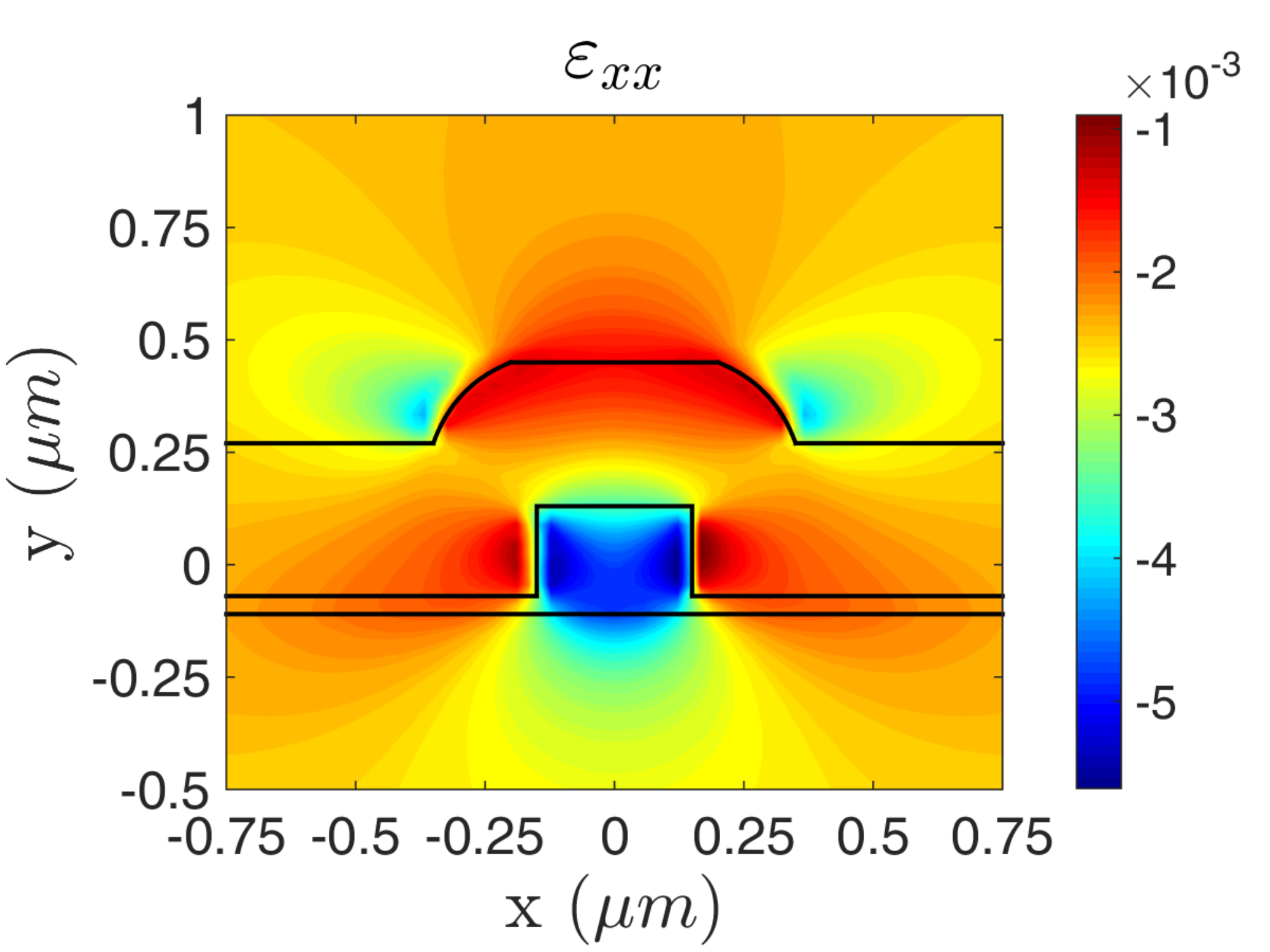}}\quad
\subfigure[\protect \label{fig:eyy_senza_crac}]{\includegraphics[width=0.4\columnwidth]{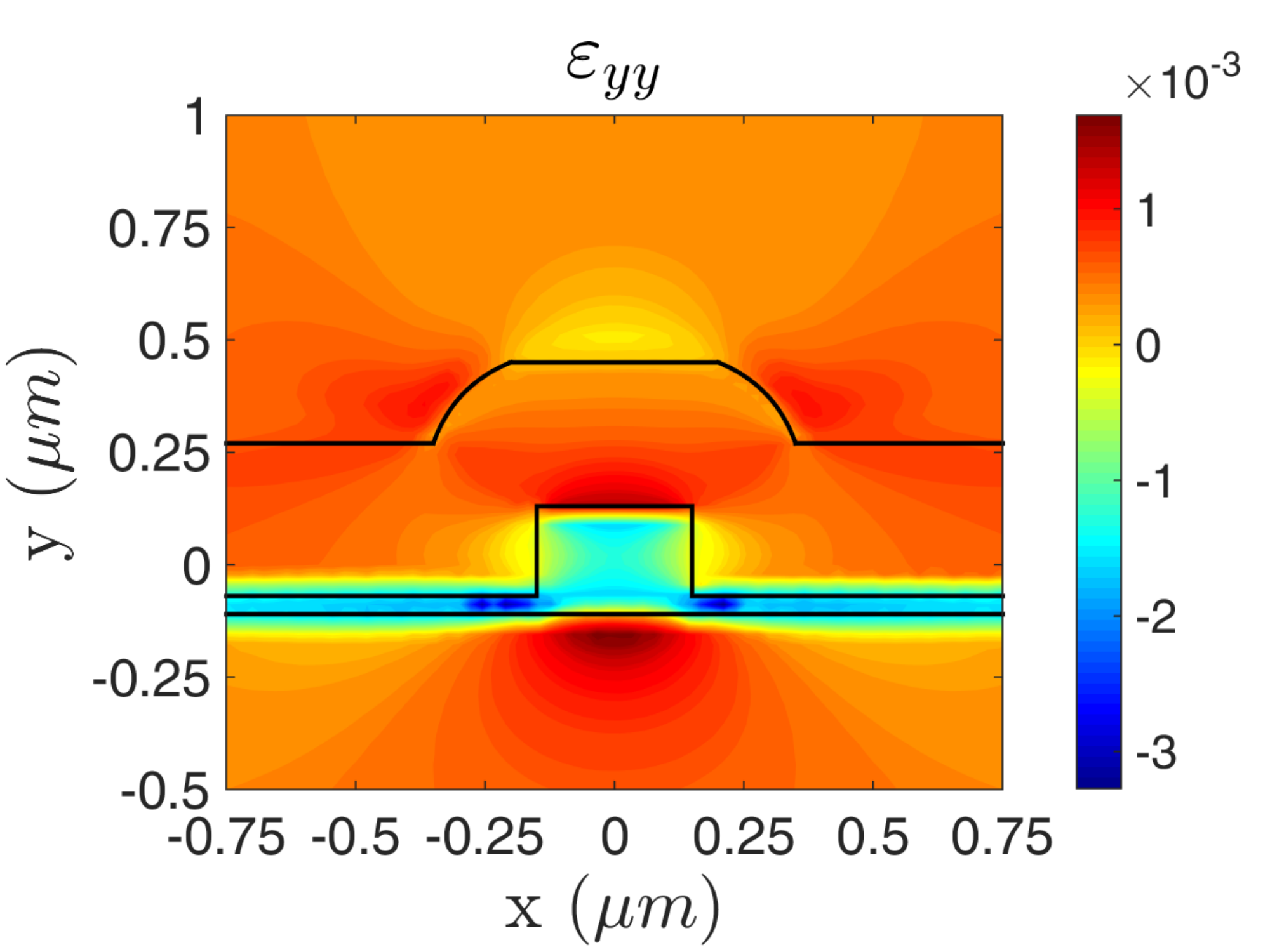}}\\
\subfigure[\protect \label{fig:ezz_senza_crac}]{\includegraphics[width=0.4\columnwidth]{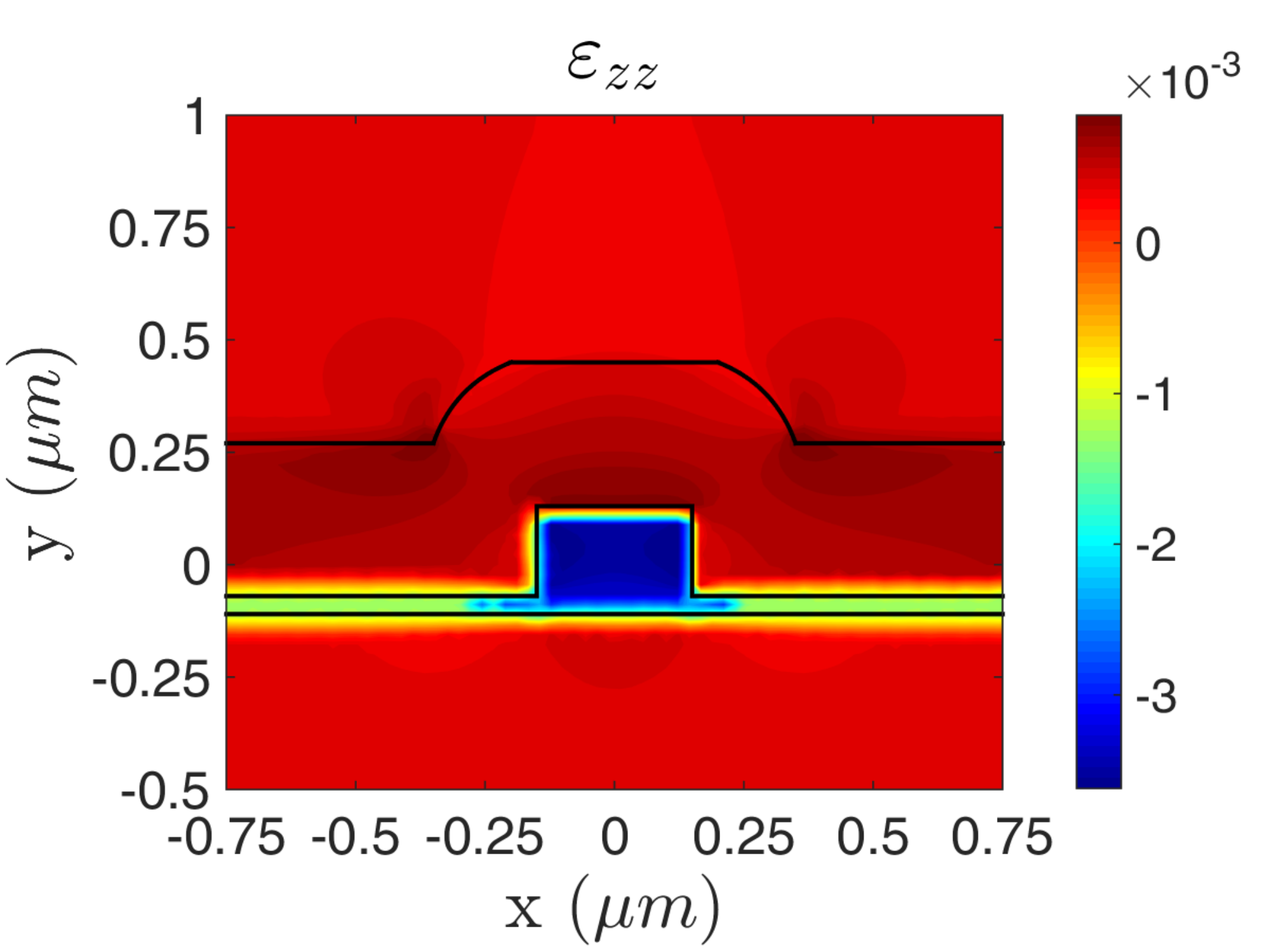}}\quad
\subfigure[\protect \label{fig:exy_senzacrac}]{\includegraphics[width=0.4\columnwidth]{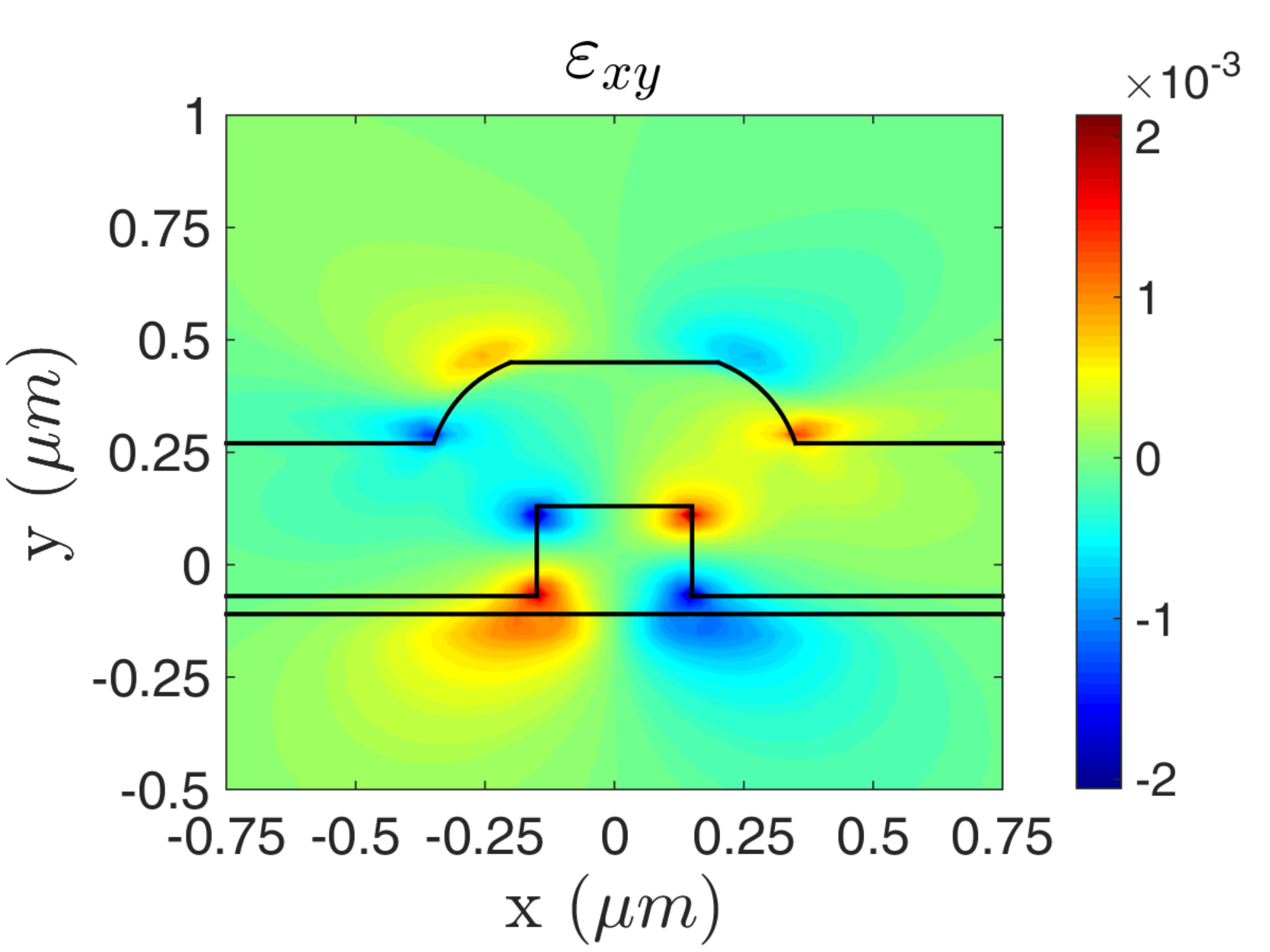}}
\caption{Strain profile of silicon waveguide used in Chmielak et al.
\cite{CHMIELAK2013}: (a) $\varepsilon_{xx}$, (b) $\varepsilon_{yy}$, (c)
$\varepsilon_{zz}$, (d) $\varepsilon_{xy}$.}
\label{fig:strain_senza_crac}
\end{figure}

\begin{figure}[!ht]
\centering
\subfigure[\protect \label{fig:Ex_field_senza_crac}]{\includegraphics[width=0.325\columnwidth]{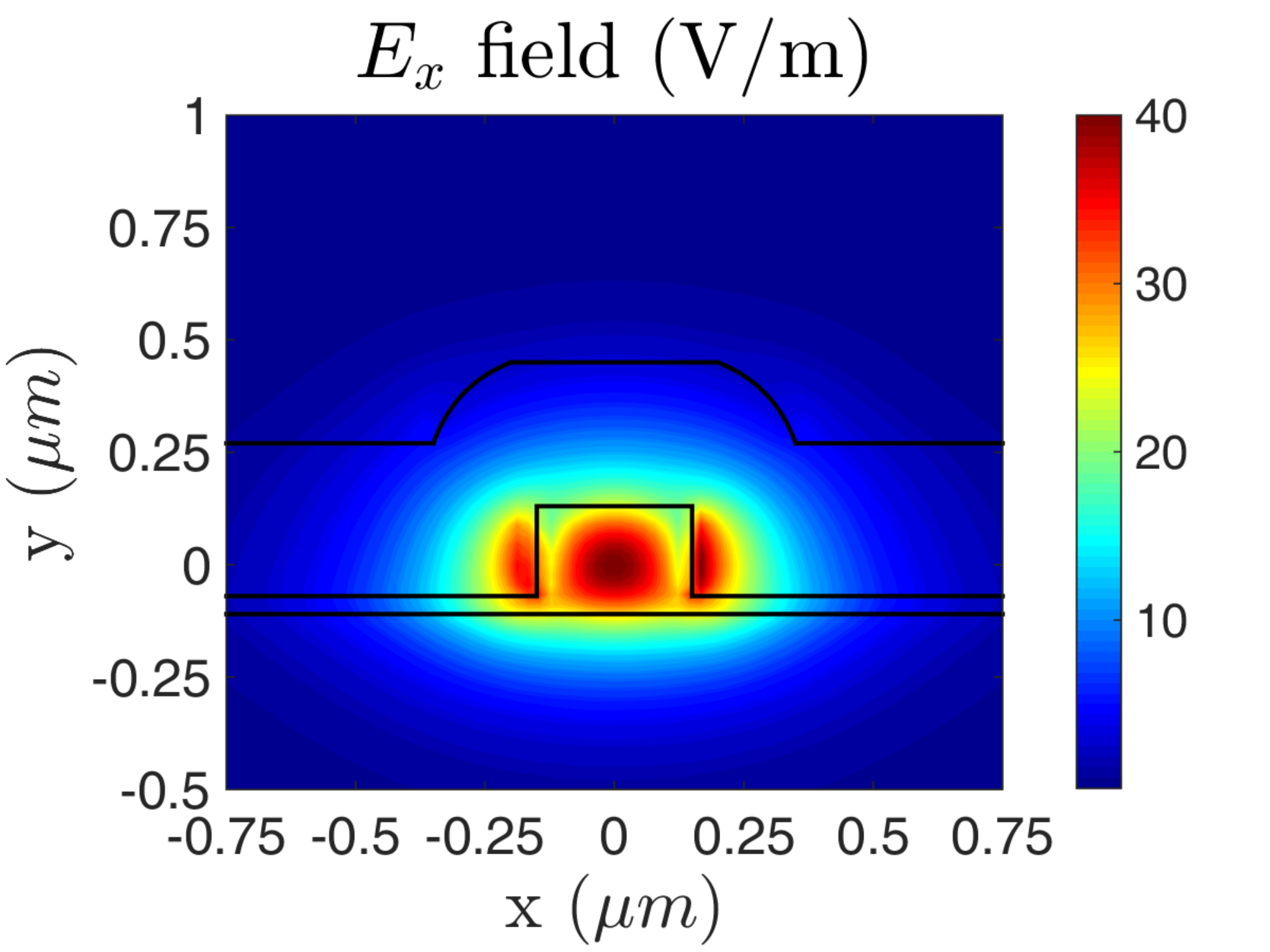}}
\subfigure[\protect \label{fig:Ey_field_senza_crac}]{\includegraphics[width=0.325\columnwidth]{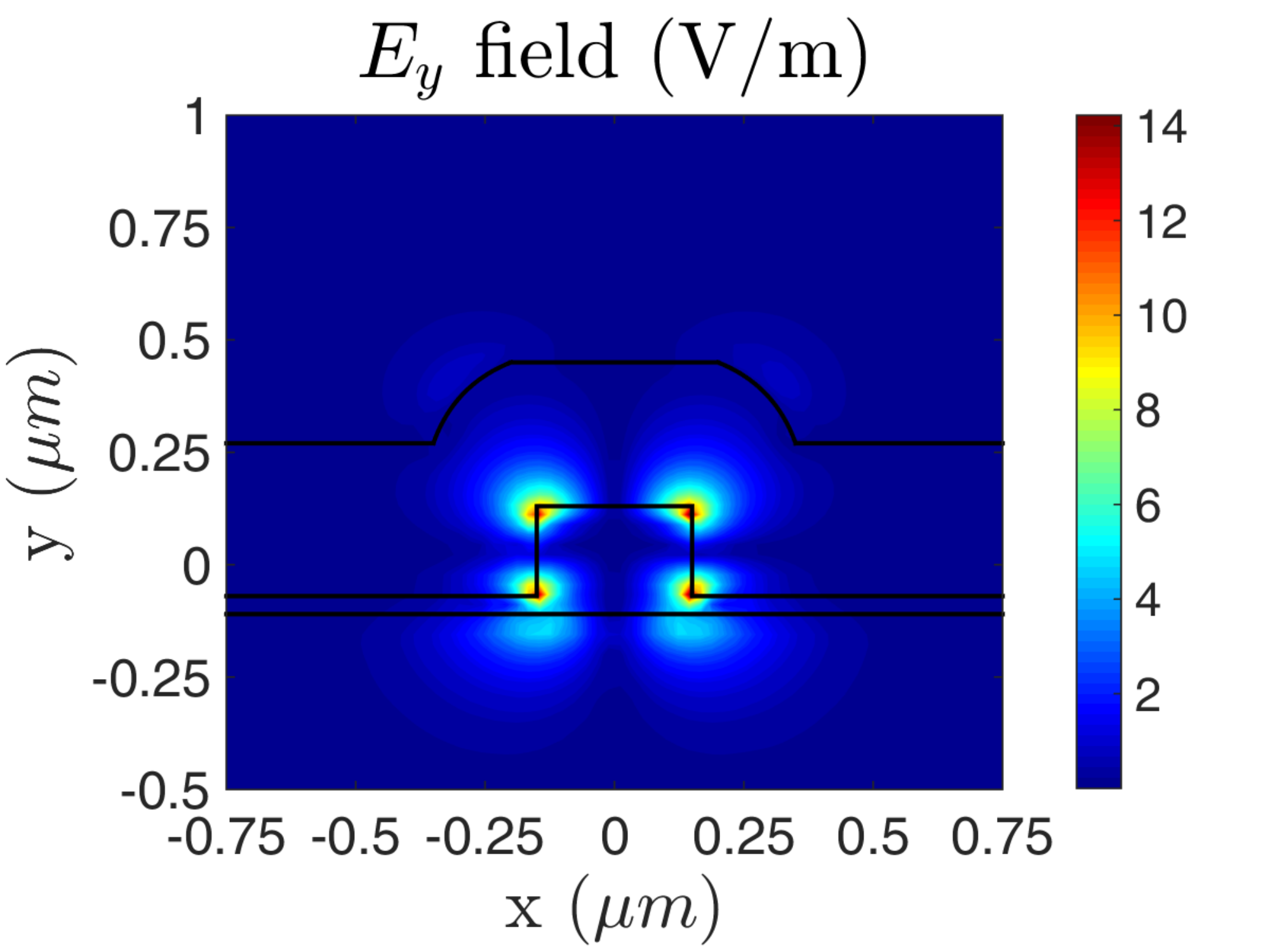}}
\subfigure[\protect \label{fig:Ez_field_senza_crac}]{\includegraphics[width=0.325\columnwidth]{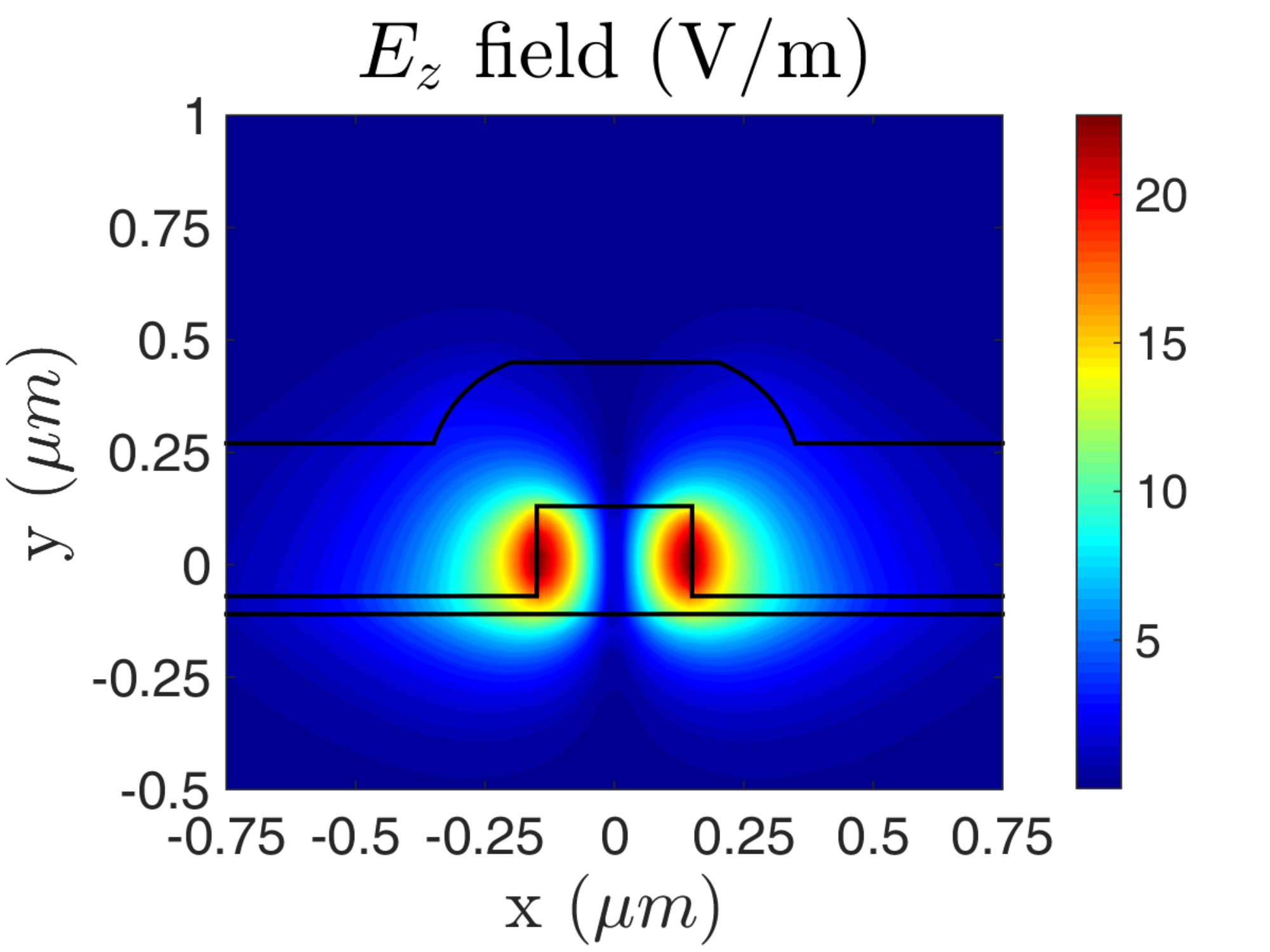}}
\caption{Electric field components in the case of Chmielak et al. \cite{CHMIELAK2013}
}\label{Chmielak_modeAnalysis}
\end{figure}

\begin{figure}[!ht]
\centering
\subfigure[\protect \label{fig:exx_pedro}]{\includegraphics[width=0.4\columnwidth]{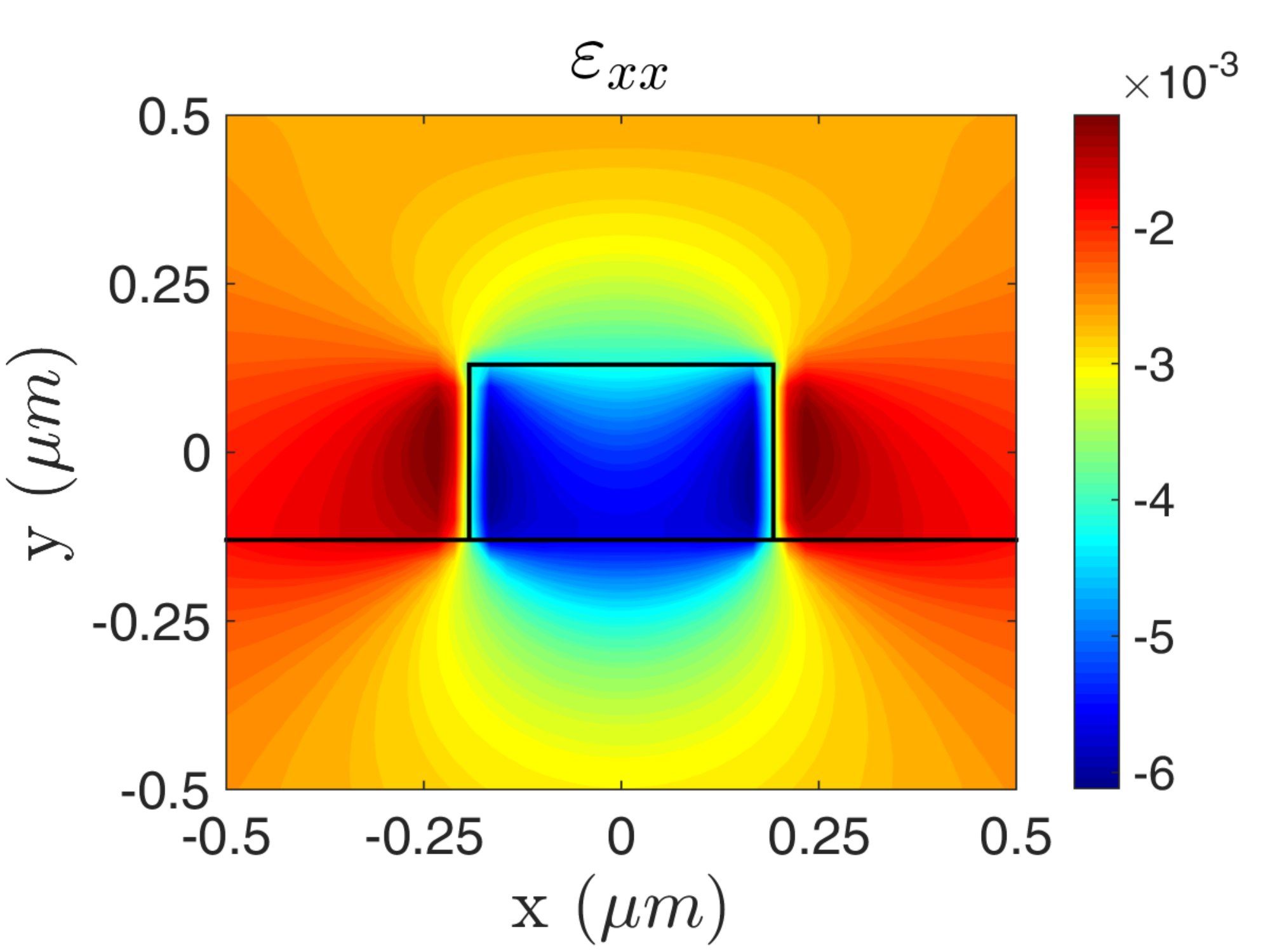}}\quad
\subfigure[\protect \label{fig:eyy_pedro}]{\includegraphics[width=0.4\columnwidth]{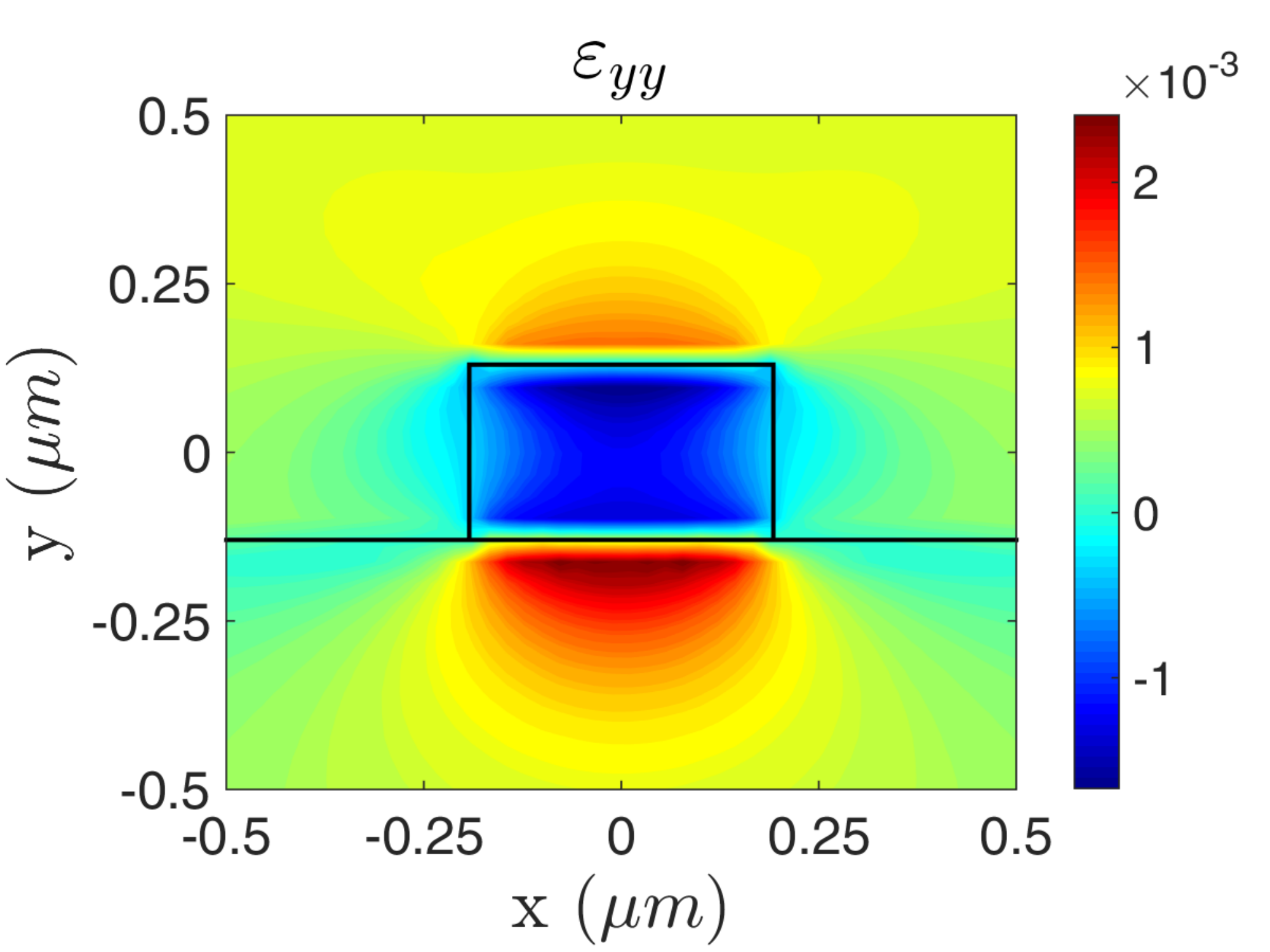}}\\
\subfigure[\protect \label{fig:ezz_pedro}]{\includegraphics[width=0.4\columnwidth]{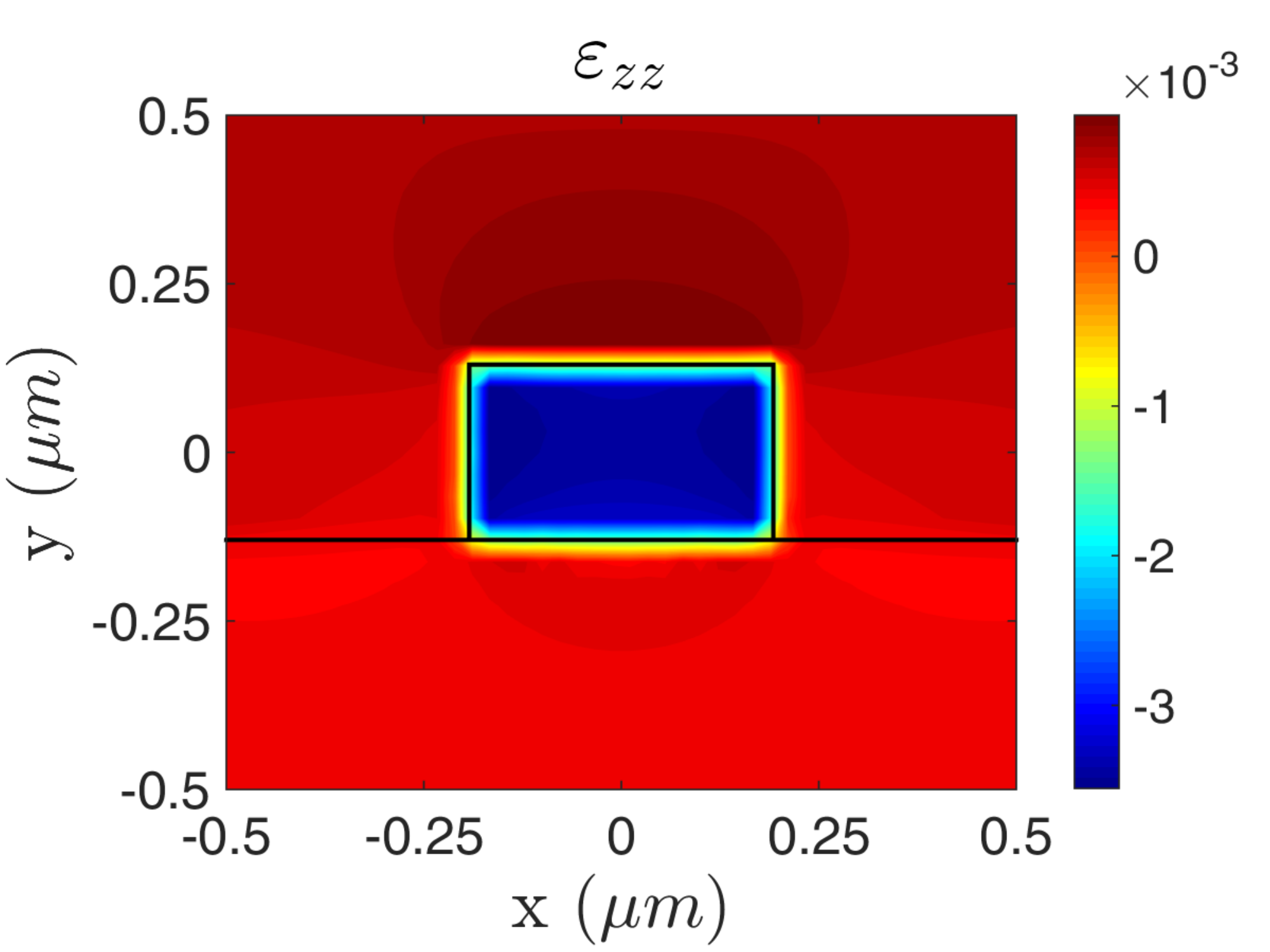}}\quad
\subfigure[\protect \label{fig:exy_pedro}]{\includegraphics[width=0.4\columnwidth]{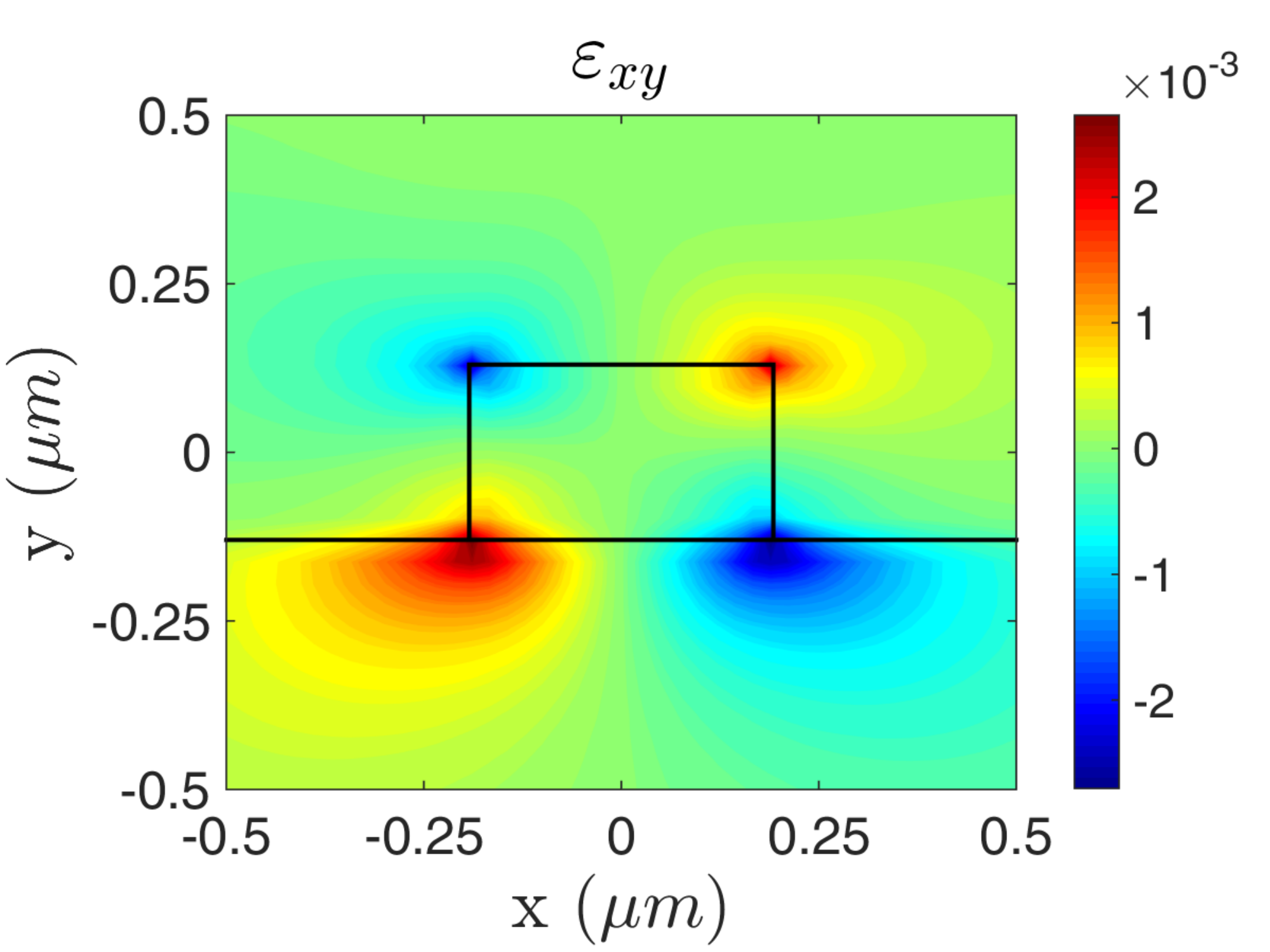}}
\caption{Strain profile of silicon waveguide used in Damas et al.
\cite{DAMAS2014}: (a) $\varepsilon_{xx}$, (b) $\varepsilon_{yy}$, (c)
$\varepsilon_{zz}$, (d) $\varepsilon_{xy}$.}
\label{fig:strain_pedro}
\end{figure}

\begin{figure}[!ht]
\centering
\subfigure[\protect \label{fig:campo_x_pedro_1550}]{\includegraphics[width=0.325\columnwidth]{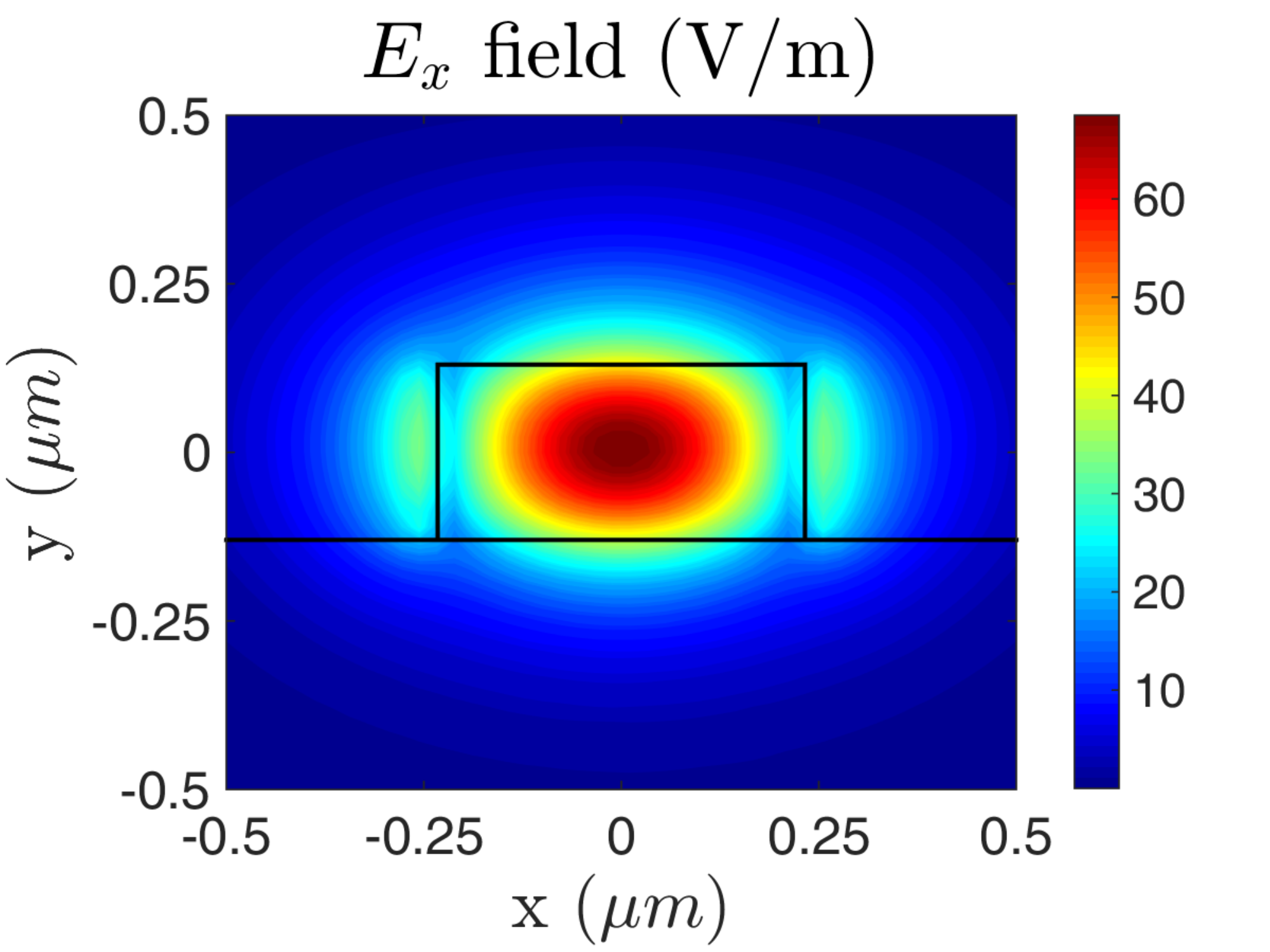}}
\subfigure[\protect \label{fig:campo_y_pedro_1550}]{\includegraphics[width=0.325\columnwidth]{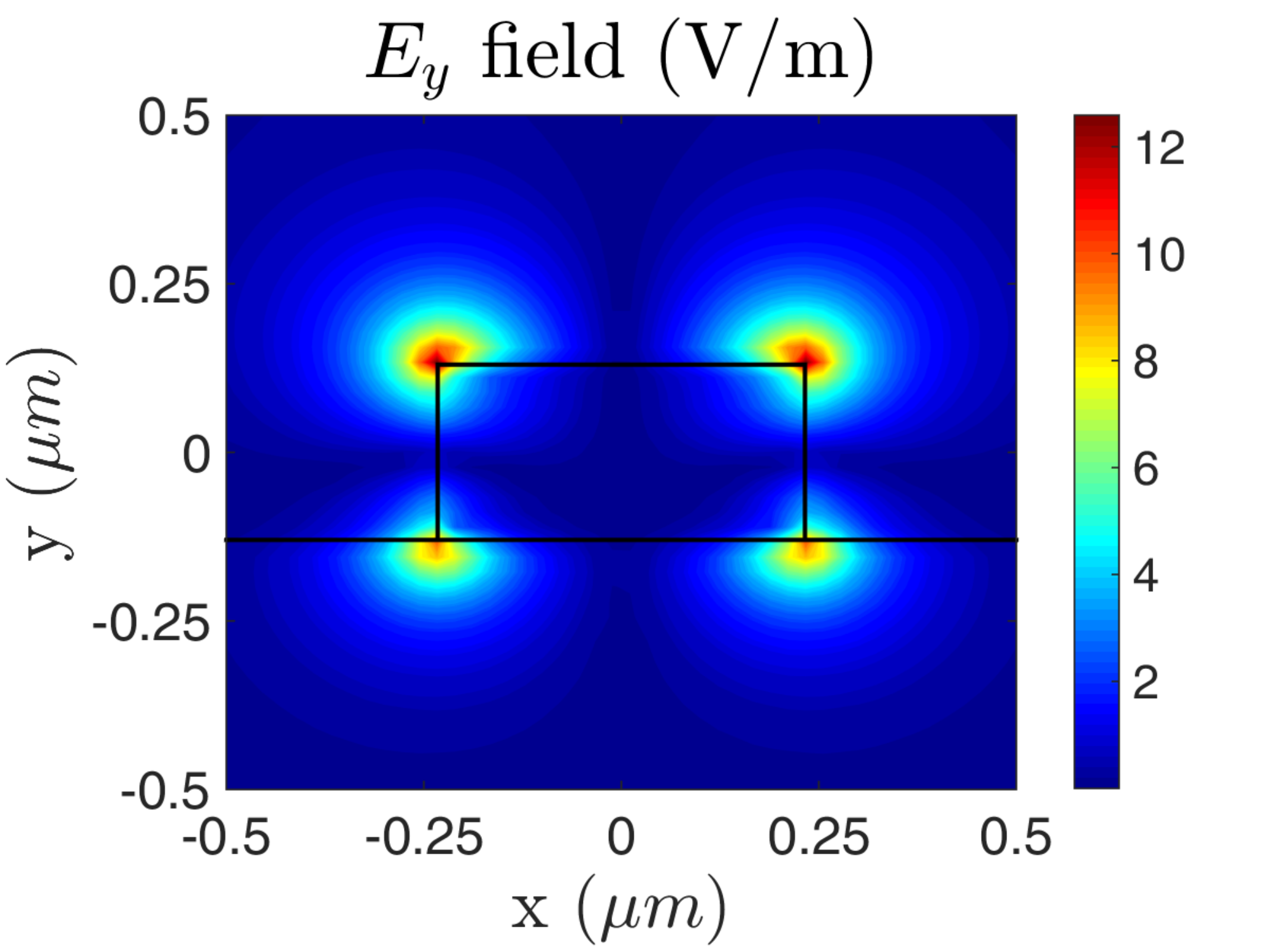}}
\subfigure[\protect \label{fig:campo_z_pedro_1550}]{\includegraphics[width=0.325\columnwidth]{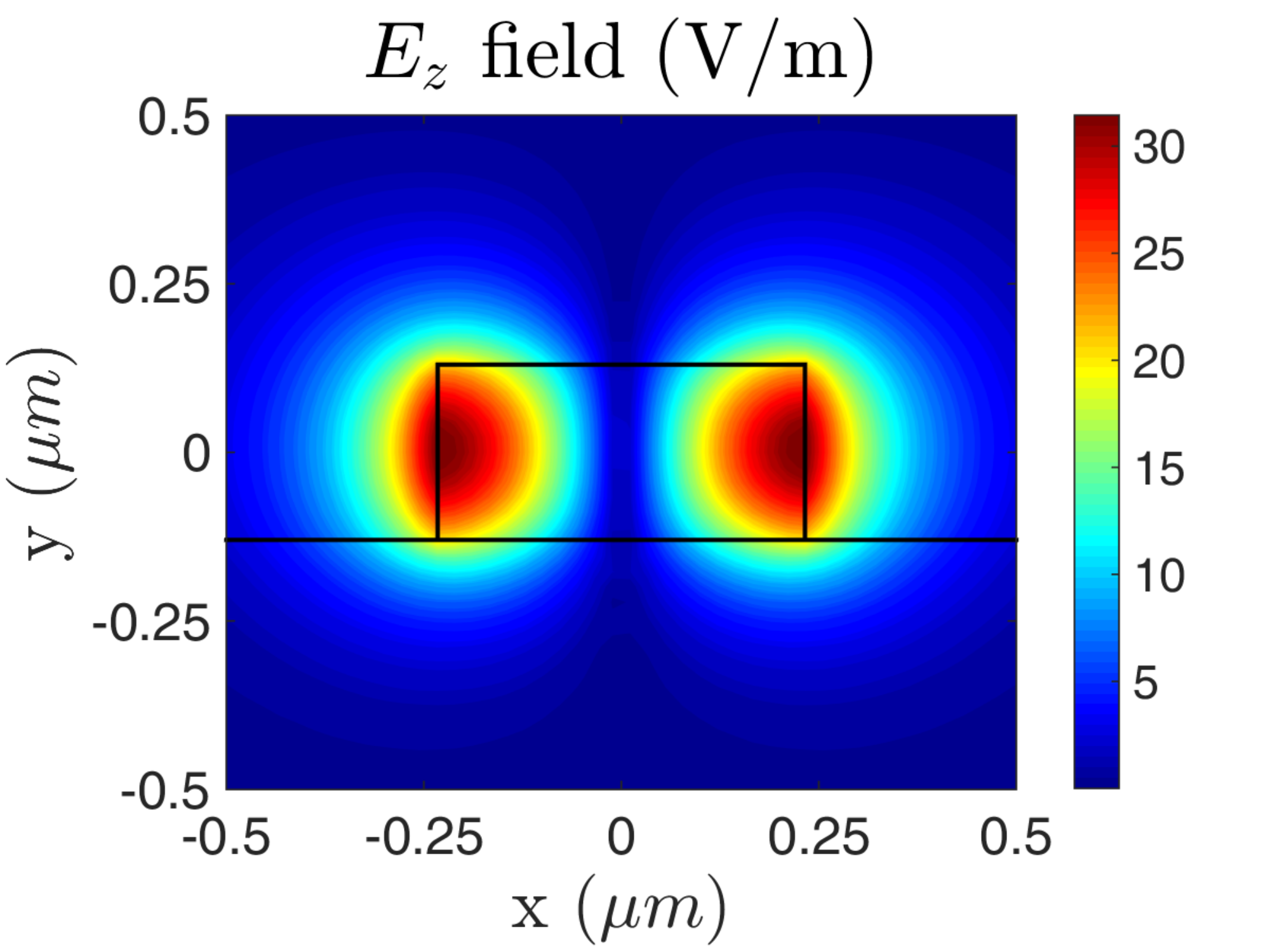}}\\
\caption{Electric field components in the case of Damas et al.~\cite{DAMAS2014}
for an incident wavelength of 1550 nm.}\label{Pedro_modeAnalysis}
\end{figure}

\subsection{Results}
In Fig.~\ref{fig:strain_senza_crac} we show the behavior of the strain
components from which we can evaluate the strain  derivatives, in
Fig.~\ref{Chmielak_modeAnalysis} the results of the electromagnetic mode
analysis, performed using COMSOL multi-physics  tools~\cite{COMSOL}. The
geometry taken into account is the waveguide cross section for the device
described in \cite{CHMIELAK2013}.  The same results for the device described in
\cite{DAMAS2014} are respectively shown in Figs.~\ref{fig:strain_pedro} and
\ref{Pedro_modeAnalysis}. Focusing on the mode analysis, it is worth noting
that in both cases the waveguides show a single mode behavior, however the
$E_y$ and $E_z$ components are not negligible compared to the $E_x$ component
and thus the mode is not purely transverse electric. The high value of $E_z$ is
due to the high index step of the waveguides which requires in fact a full
vectorial mode solver to be accurately computed~\cite{RAHMAN1984}.

Working in the device coordinate frame reported in Figs.~\ref{fig:Chmielak} and
\ref{fig:Damas} and with $\bm{E}^{dc}$ directed along the $y$-axis, the
relation between the effective susceptibility and the strain gradient can be
written in the form 
\begin{equation}\label{eq:effectiveSuscpetibility_ourCase}
\chi^{\mathrm{eff}}_y(\omega)=c_i o_i(\omega)\ ,
\end{equation}
where the coefficients $c_i$ ($i=1,\ldots,15$) are the independent entries of
the tensor $\bm{T}$ and the terms $o_i$ are linear combinations of the weighted
strain gradients that will be called overlap functions in the following.  Their
explicit form is reported in App.~\ref{sec:explicit_form},
Eq.~\eqref{eq:overlapFactorDef} and it is the same for both the
crystallographic axis orientation considered in \cite{CHMIELAK2011,
CHMIELAK2013} and \cite{DAMAS2014}. For these orientations some of the overlap
factors turn out to coincide:
\begin{equation}\label{eq:overlapFactorIdentities}
\begin{aligned}
o_{3p7}\equiv o_{3}=o_{7},\qquad
o_{5p6}\equiv o_{5}=o_{6},\qquad
o_{11p12}\equiv o_{11}=o_{12},
\end{aligned}
\end{equation}
so that only $12$ independent coefficients are needed in this case, instead of the
$15$ ones required for a generic device orientation.

\begin{figure}[!t]
\centering
\subfigure[]{
\includegraphics[width=0.45\columnwidth]{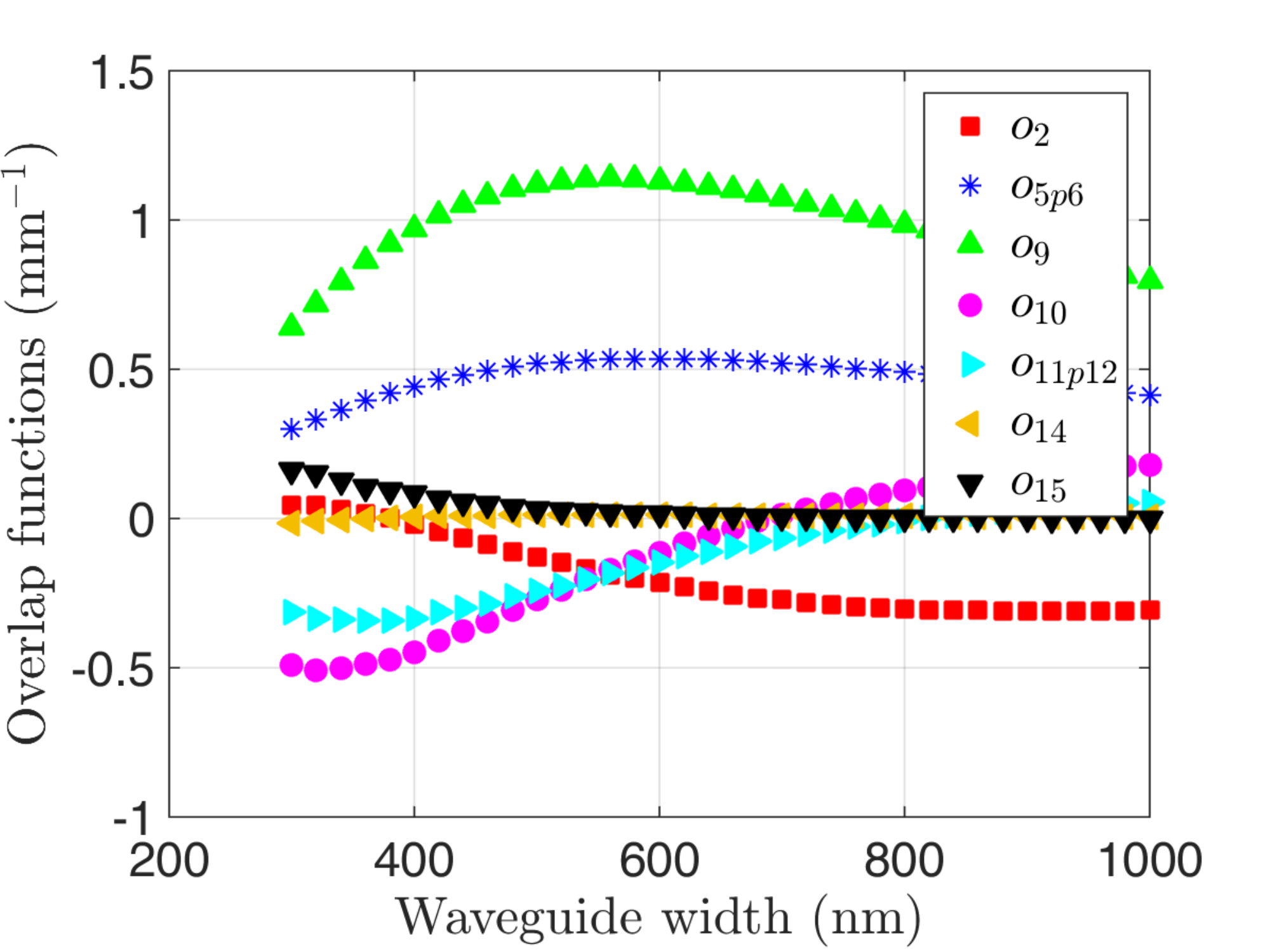}} 
\quad
\subfigure[]{
\includegraphics[width=0.45\columnwidth]{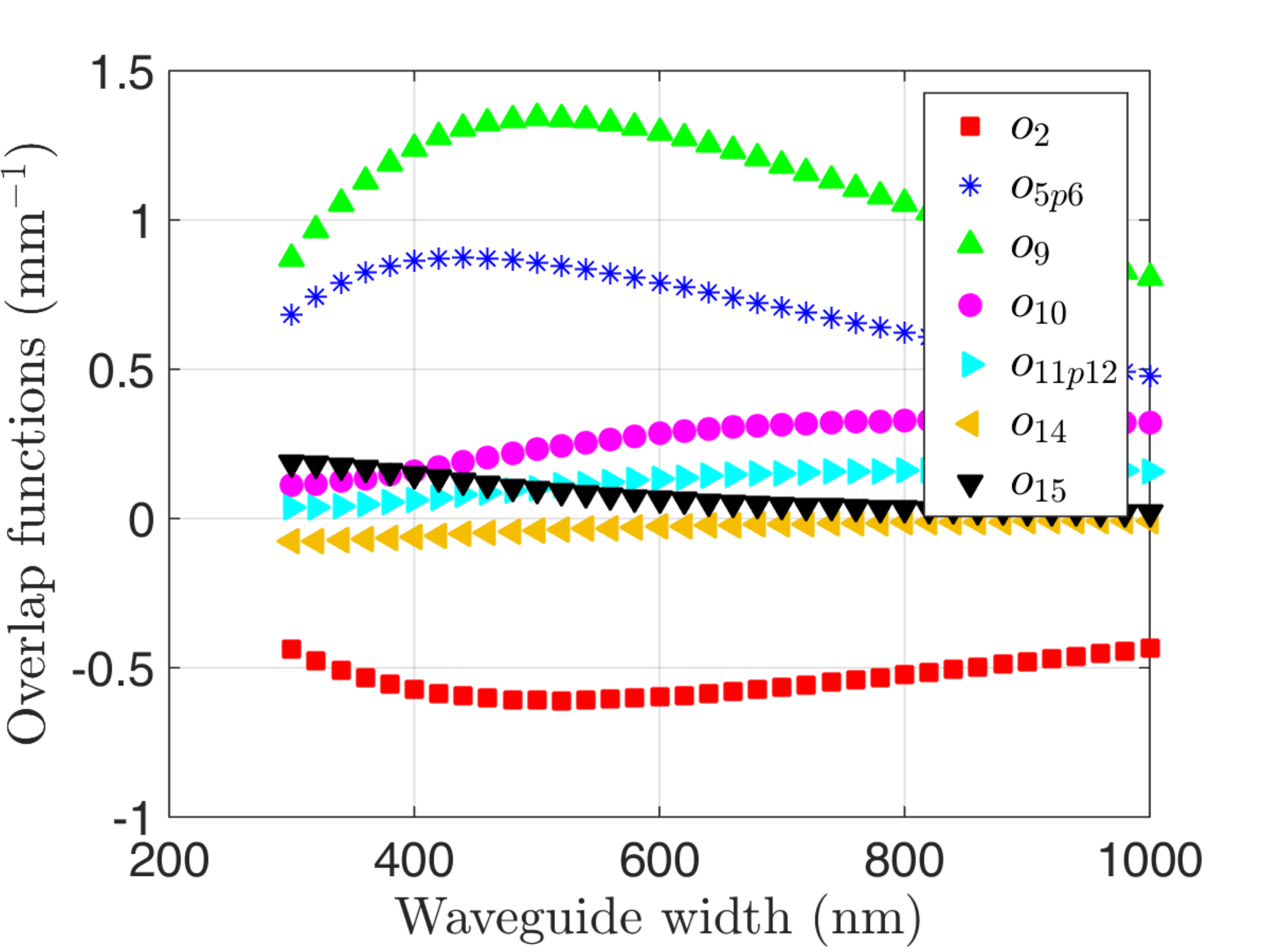}}
\caption{Behavior of the overlap functions for the waveguides under
investigation with respect to waveguide width for (a) the device used by
Chmielak et al.~\cite{CHMIELAK2013} and (b) the device used by  Damas et
al.~\cite{DAMAS2014}. In both cases only the most significant overlaps are
plotted at $\lambda=1550nm$.} \label{overlap_wrt_width}
\end{figure}

\begin{figure}[!t]
\centering
\subfigure[]{
\includegraphics[width=0.45\columnwidth]{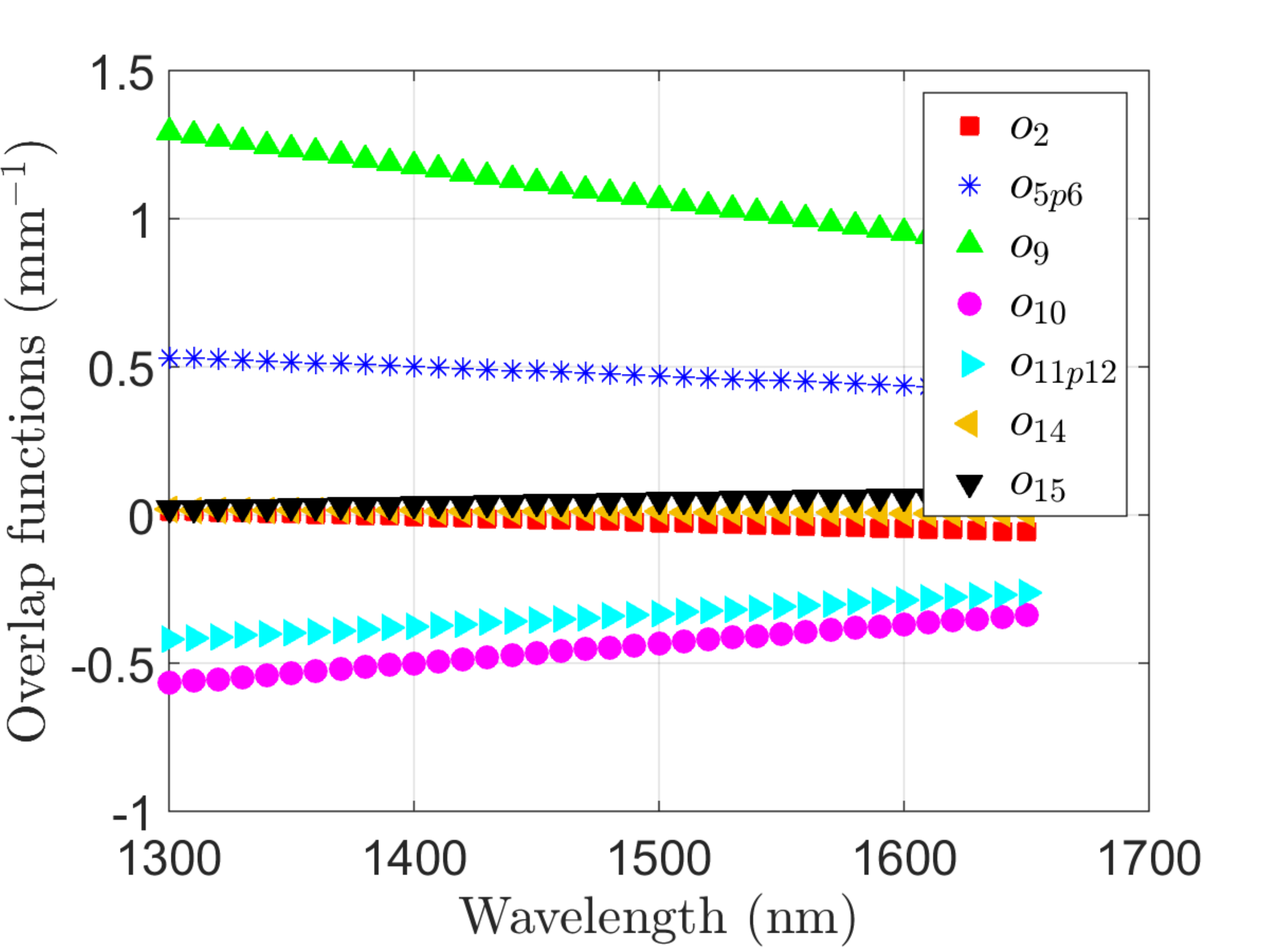}} 
\quad
\subfigure[]{
\includegraphics[width=0.45\columnwidth]{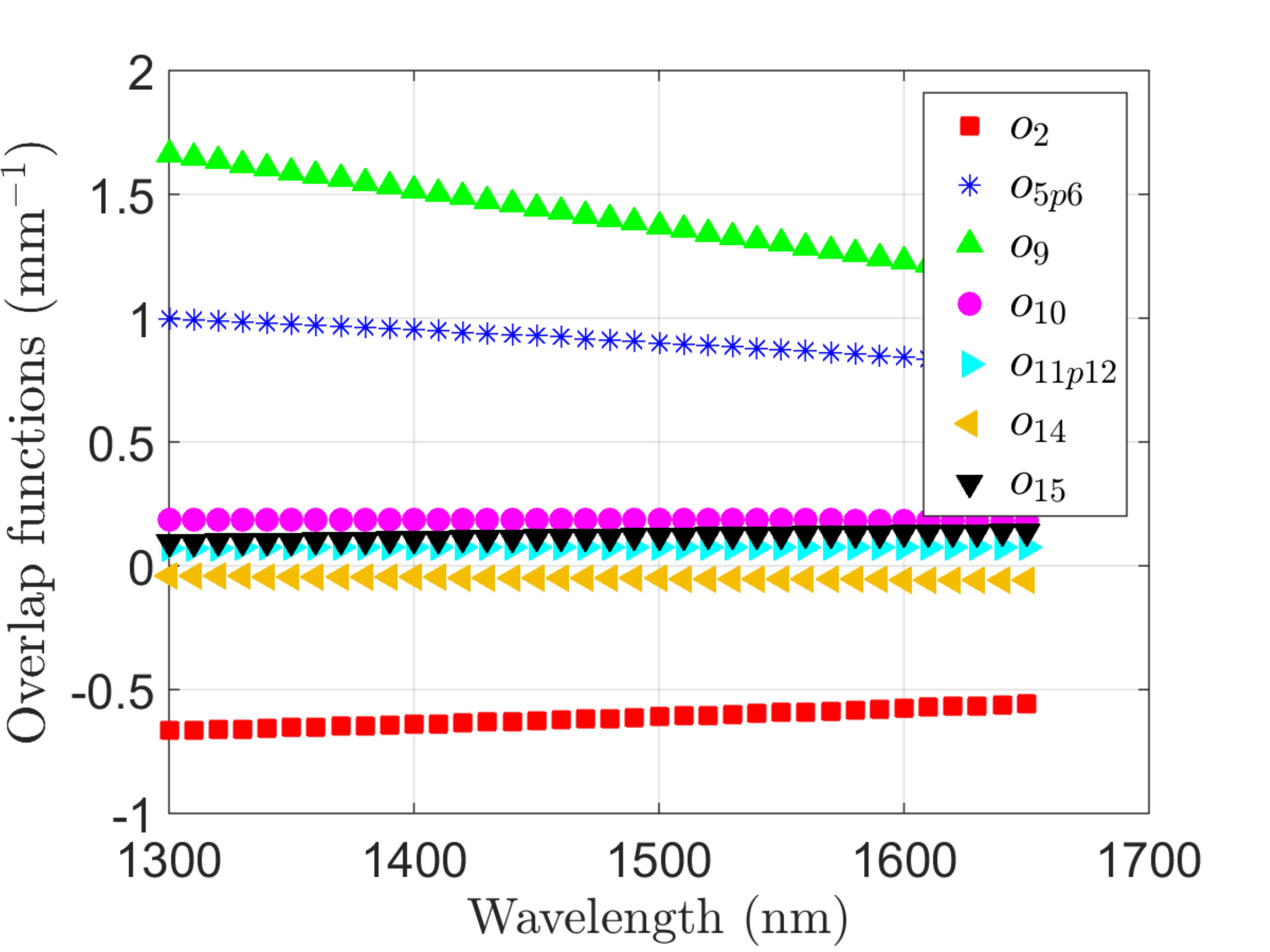}}
\caption{Behavior of the overlap functions for (a) the waveguide used in Chmielak et al.~\cite{CHMIELAK2013} and (b) the waveguide used in Damas et al.~\cite{DAMAS2014} with respect to the wavelength for a waveguide width $w_{Si}=385\,nm$. Only the most significant overlaps are plotted.
\label{overlap_wrt_lambda_Pedro}}
\end{figure}

It is reasonable to expect that, for practical purposes, the number of
numerical constants to be fixed can be further reduced. Indeed we numerically
estimated the overlap functions corresponding to the experimental setups used
in \cite{CHMIELAK2013} and \cite{DAMAS2014} and observed that in both cases
a clear hierarchy can be observed: the values of $o_{2}$, $o_{5p6}$, $o_{9}$,
$o_{10}$ and $o_{11p12}$ are significantly larger than the others, with
marginal contributions from $o_{14}$ and $o_{15}$; all the other overlaps are
smaller by an order of magnitude or more. In Figs.~\ref{overlap_wrt_width} and
Fig.~\ref{overlap_wrt_lambda_Pedro} these most significants overlap are
displayed for various waveguide width and wavelength.  Since all the symmetries
of the problem have been taken into account when determining the independent
components of the tensor $\bm{T}$, it is natural to expect all the $c_i$ to be
about the same order of magnitude.  As a consequence one expects that the
coefficients that will be more important to reliably describe experimental
results are the ones multiplying the dominant overlaps. This reduces the number
of independent constants to be determined to $7$ in our more conservative
estimate.

From Figs.~\ref{overlap_wrt_width} and \ref{overlap_wrt_lambda_Pedro} we see
that the dominant contributions are the ones denoted by $o_9$, $o_{5p6}$ and
$o_2$.  It is thus interesting to look at the specific form of these overlaps
in Eq.~\eqref{eq:overlapFactorDef}. The largest field component is $E_x$ and it
is natural to expect that these large overlaps are weighted by $E_x$, which is
indeed the case. Regarding the strain-gradient components, $o_{5p6}$ and $o_9$
involve $\partial\varepsilon_{xx}/\partial y$, which has been
often assumed in the literature as the only relevant component of the
strain-gradient; however $o_2$ is independent of this component and is related
to $\partial\varepsilon_{yy}/\partial y$, whose importance has been overlooked so far.

\begin{figure}[!ht]
\centering
\subfigure[\protect \label{fig:exx_mezzocrac}]{\includegraphics[width=0.4\textwidth]{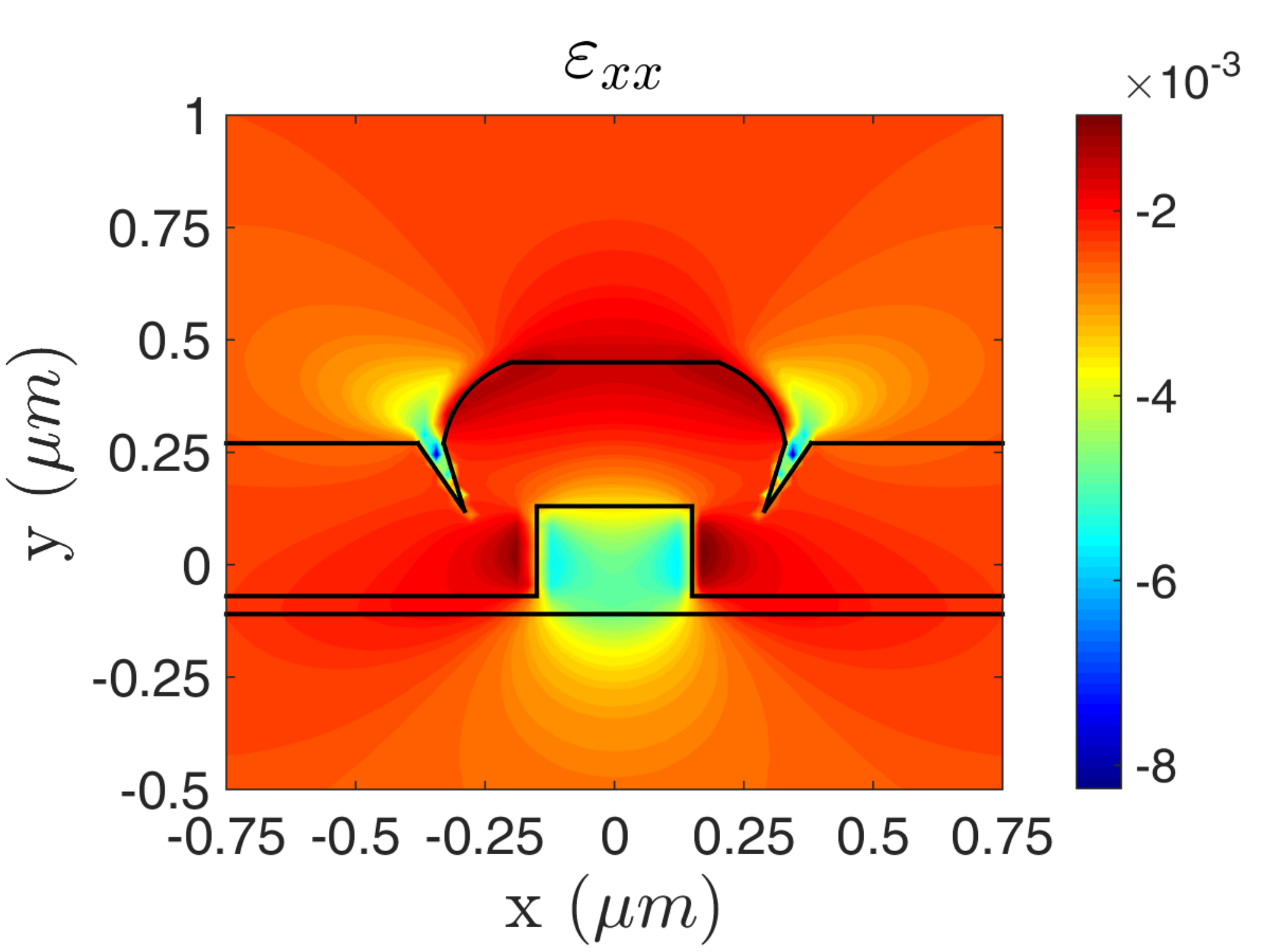}}
\subfigure[\protect \label{fig:eyy_mezzocrac}]{\includegraphics[width=0.4\textwidth]{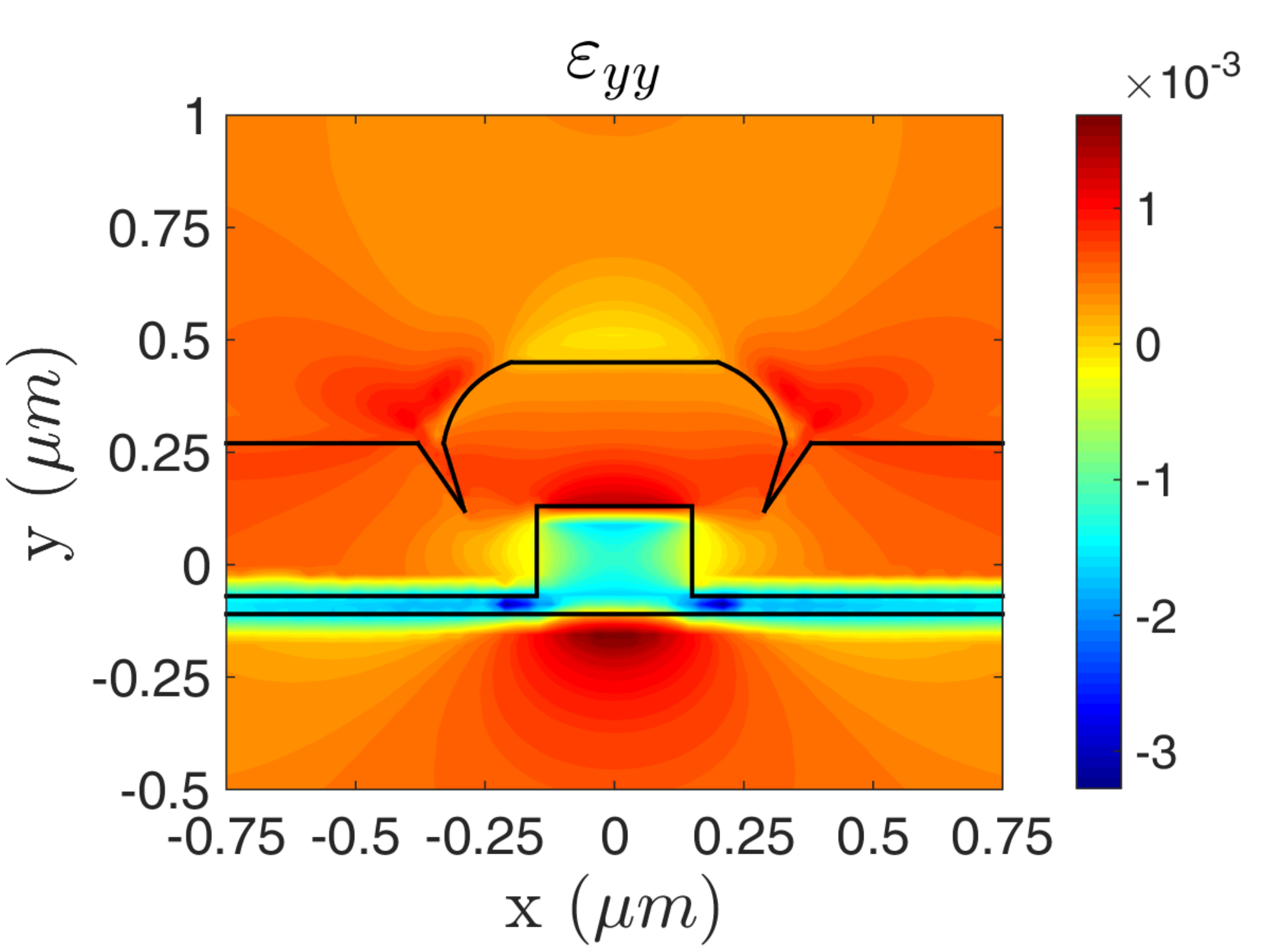}}\\
\subfigure[\protect \label{fig:ezz_mezzocrac}]{\includegraphics[width=0.4\textwidth]{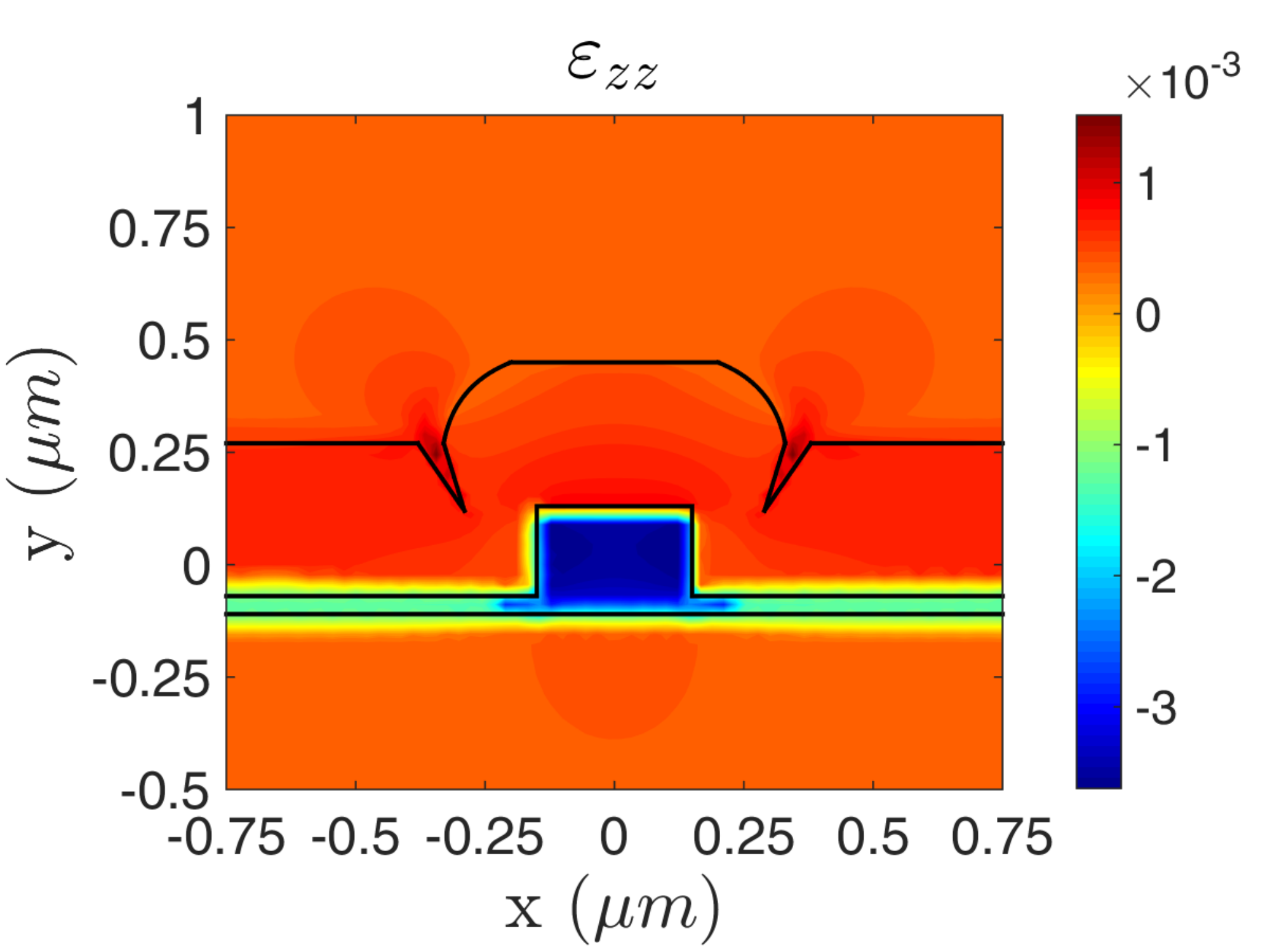}}
\subfigure[\protect \label{fig:exy_mezzocrac}]{\includegraphics[width=0.4\textwidth]{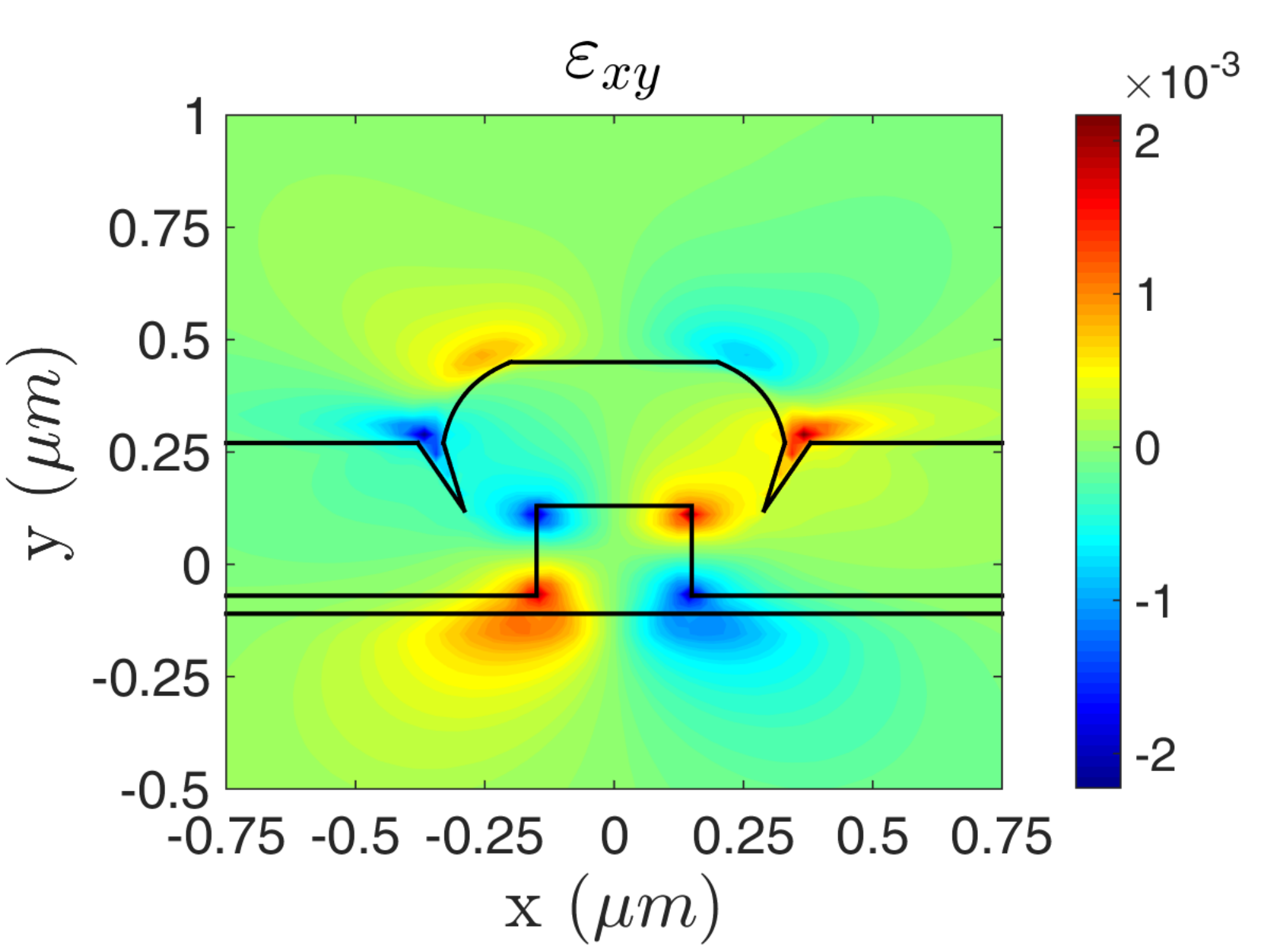}}
\caption{Strain components profile $\varepsilon_{xx}$ (a),
$\varepsilon_{yy}$(b), $\varepsilon_{zz}$(c) and $\varepsilon_{xy}$(d) in the
strained silicon waveguide \cite{CHMIELAK2013} in presence of defect fabrication in the Si$_3$N$_4$
slab.}
\label{fig:strain_mezzocrac}
\end{figure}

\begin{figure}[!th]
\centering{
\includegraphics[width=0.45\columnwidth]{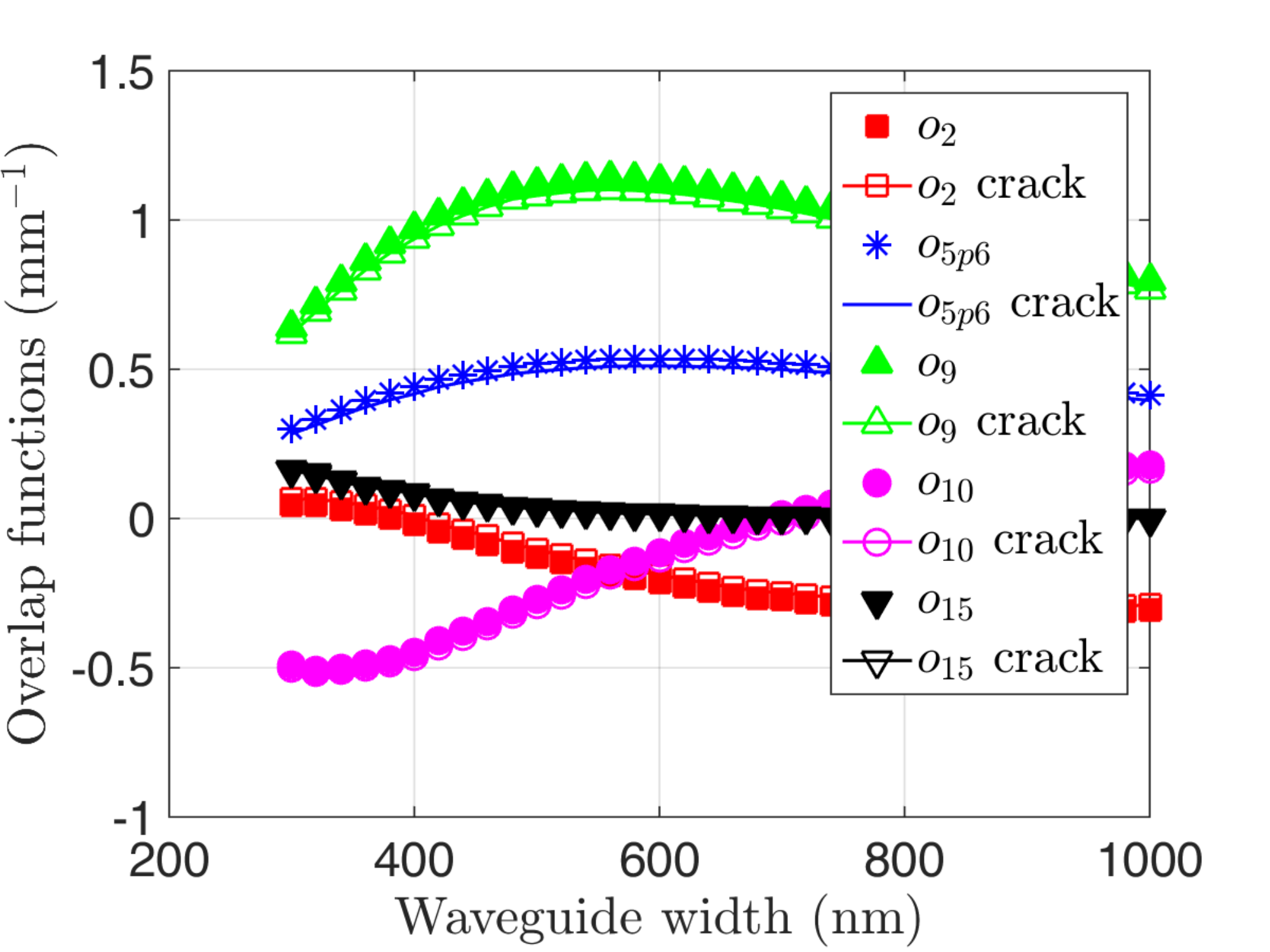}
\caption{Comparison between the overlap factors for the case of Chmielak et
al.~\cite{CHMIELAK2013} with and without fabrication defect.}
\label{overlap_confronto_consenzacrac}}
\end{figure}

As explained in the introduction, we cannot at present estimate the numerical
values of the constants $c_i$ in Eq.~\eqref{eq:effectiveSuscpetibility_ourCase}
for lack of reliable (i.e. for which we can safely exclude a contamination from
free carriers effects) experimental data to compare with. It is however
interesting to note that, also without any knowledge of these constants, we can
reach an important conclusion on a point that attracted some attention in the
literature, i.e. the relevance of fabrication defects on the strain-induced
Pockels effect.  
In particular in \cite{CHMIELAK2013} was noted that the strain profile of the
studied structure was quite sensitive to the presence of lateral cracks
in the $\mathrm{Si}_3\mathrm{N}_4$ overlayer.  
When such defects are present, we observe a strong increase of both
strain components and strain gradients in proximity of the lateral cracks, as
shown in Fig.~\ref{fig:strain_mezzocrac}. 
However, as can be seen from
Fig.~\ref{overlap_confronto_consenzacrac}, the defects do not induce any
sizable modification of the overlap functions. 

The physical explanation for this result is that the device
regions in which the strain is significantly affected by the defect are also
the regions where the electromagnetic field is weak, so that the net effect of
the crack on the quantities in Eq.~\eqref{eq:weighted_zeta} is small.  This
simple fact was previously overlooked, probably due a confusion that is
sometimes present in the literature between the strain-gradient dependence of
the local non-linear susceptibility in Eq.~\eqref{eq:T_definition} and that of
the effective susceptibility, in which the strain-gradient tensor enters
weighted by the electromagnetic field.  Having explicitly checked that the
overlaps does not appreciably changes when the lateral crack is present, we can
thus safely conclude (also without any knowledge of the numerical coefficient
$c_i$) that the dependence of the effective susceptibility on this type of
fabrication defects is very weak, as far as the strain-induced Pockels effect
is concerned.

\section{Conclusions}\label{sec:concl}

We have proposed an effective model that describes the
strain-induced dielectric susceptibility in centro-symmetric crystals starting
from symmetry arguments. The specific case of the Pockels effect in strained
silicon has been investigated as a particularly interesting example, with
potential applications in silicon photonics for implementing modulation and
switching functions, however the idea of the approach is extemely general and
can be applied to any centro-symmetric crystals (in fact also to
non centro-symmetric ones) and to other physical phenomena, like e.g. second
harmonic generation.

The main result of our analysis consists in a simple relation between the
second order dielectric susceptibility and the strain
gradient tensor.  We have shown that the effective second order susceptibility
of strained silicon cannot be accurately described by considering only
mechanical deformations but the combination of optical modes and mechanical
stress analysis is required. As a result the effective susceptibility can be
written as a linear combination of the weighted strain gradient components
(defined in Eq.~\eqref{eq:weighted_zeta}) with fifteen independent
coefficients in the general case. This approach put on firm ground 
some intuitive ideas present in the recent literature, that suggested  a relation
between the strain-induced effective susceptibility and the strain gradient.

In the experimental settings studied in \cite{CHMIELAK2011,CHMIELAK2013}
and \cite{DAMAS2014} the number of independent coefficients to be determined to
completely parametrize the strain-induced Pockels effect reduces to $12$,
moreover we gave indications that, for all practical purposes, this number can
be effectively reduced to $7$.

When new experimental measures of the effective strain-induced
susceptibility (free from spurious effects like the ones related to free
carriers) will be available, it will be possible to estimate these
coefficients.This
will have a considerable impact on the design and optimization of
electro-optic modulators based on strained silicon, since it reduces the
computation of electro-optic effect to standard mechanical and optical
computations, thus allowing device
optimization in terms of silicon waveguide geometries, crystallographic axes
orientation and electrode position.

\appendix

\section{Contracted index notation}\label{sec:compacted}

When dealing with tensors which are invariant under permutation of some indices
it is convenient to introduce a compact notation, in order to simplify
expressions. It is customary to define the contracted index notation (denoted
by curly brackets) as follow 
\begin{equation}
\begin{aligned}
& \{11\}\rightarrow 1, & & \{22\}\rightarrow 2, & & \{33\}\rightarrow 3, \\
& \{23\},\{32\}\rightarrow 4, & & \{13\},\{31\}\rightarrow 5, & & \{12\},\{21\}\rightarrow 6.
\end{aligned}
\end{equation}
Using this contracted notation a symmetric $2-$index tensor $\bm{V}$ (i.e.,
$V_{ij}=V_{ji}$) can be written as a vector $\hat{\bm{V}}$ with $6$ entries,
whose components are $V_{\{ij\}}$ with $i\le j$, and analogous simplifications
occurs also for higher order tensors. It is important to note that, while the
index contraction make some manipulations easier, the tensor transformation and
multiplication properties are changed by this index replacement and some care
is required. 

In contracted notation, the $6$-index tensor $\bm{T}$ tensor introduced in 
Eq.~\eqref{eq:T_definition} can be written (for the case of the Pockels effect,
see the end of Sec.~\ref{sec:model}) as the the four index tensor 
\begin{equation}
\hat{T}_{i_1 i_2 i_3 i_4}=T_{\{j_1 j_2\} j_3 \{j_4 j_5\} j_6}\ ,
\end{equation}
where $j$-indices take value in $1,2,3$, while $i_2,i_4=1,2,3$, and $i_1,i_3=1,\ldots,6$.

\section{Symmetry analysis of the tensor {\it T}}\label{sec:TSymmetriesAlg}

In this appendix we report the algorithm used to write down the independent
components of the $6-$index tensor $\bm{T}$ introduced in
Eq.~\eqref{eq:T_definition}. It is first of all convenient to rewrite the
tensor $\bm{T}$ as a column vector $\bm{T}^c$ by introducing a lexicographic
labelling of the indices. The component $T_{i_1 i_2\ldots i_6}$
($i_1,\ldots,i_6=1,2,3$) corresponds to the component $T_i^c$ where 
\begin{equation}\label{eq:tensor2vector_indices}
i=3^{6-1}\,(i_6-1)+\ldots+3\,(i_2-1)+i_1\ .
\end{equation}

In order to impose the invariance properties of $\bm{T}$, the general
transformation law for tensors have to be translated in a form suitable to be
applied to the vector $\bm{T}^c$. If $\bm{R}$ is the matrix representing the
coordinate transformation, let us define the square matrix $\bm{A}$ as 
\begin{equation}
A_{ij}=R_{i_1j_1}R_{i_2j_2}\ldots R_{i_6j_6}
\end{equation}
where the relation between $i$ and $i_1,\ldots,i_6$ is the same as
Eq.~\eqref{eq:tensor2vector_indices} and analogously for the $j$s. The
invariance of the tensor $\bm{T}$ under lattice symmetries is thus equivalent
to the relation $\bm{T}^c=\bm{A}\bm{T}^c$, moreover all the entries of $\bm{A}$
are integers in the crystallographic base~\cite{NYE1985}.

In the case $s$ symmetry operations are present, with associated matrices
$\bm{A}^{(i)}$, the vector $\bm{T}^c$ has to satisfy the linear system
\begin{equation}\label{compact_rot}
\begin{pmatrix}\bm{I}\\ \vdots \\ \bm{I}\end{pmatrix}\bm{T}^c
=\begin{pmatrix}\bm{A}^{(1)}\\ \vdots \\ \bm{A}^{(s)}\end{pmatrix}\bm{T}^c,
\end{equation}
where $\bm{I}$ is the identity matrix. 

To identify the independent elements of the tensor $\bm{T}$, symmetries under
index permutation must also be taken into account.  For this purpose, let us
introduce the column vector $\hat{\bm{T}}^c$ associated to the contracted
tensor $\hat{\bm{T}}$, defined by $\hat{T}_{i}^c=\hat{T}_{i_1 i_2 i_3 i_4}$,
where $i_2,i_4=1,2,3$, $i_1,i_3=1,\ldots,6$ and 
\begin{equation}\label{compactedtensor2vector}
i=6\cdot3\cdot6\cdot(i_4-1)+6\cdot 3\cdot(i_3-1)+6\cdot(i_2-1)+i_1\ .
\end{equation}
The relation between $\hat{\bm{T}}$ and $\hat{\bm{T}}^c$ can be written in matrix form as 
\begin{equation}\label{compact_sym}
\bm{T}^c=\bm{C}\hat{\bm{T}}^c\ , 
\end{equation}
where $\bm{C}$ is a $729\times 324$ matrix, whose rows have only a single
element different from zero and equal to~$1$. Combining Eq.~\eqref{compact_rot}
and Eq.~\eqref{compact_sym} we arrive to
\begin{equation}\label{finalSystem}
0=\begin{pmatrix}\bm{C}-\bm{A}^{(1)}\bm{C}\\ \vdots \\ \bm{C}-\bm{A}^{(s)}\bm{C}\end{pmatrix}
\hat{\bm{T}}^c \equiv\bm{N}\hat{\bm{T}}^c\ ,
\end{equation}
where $\bm{N}$ is a rectangular matrix.  Because $\bm{N}$ is real,
$\mathrm{rank}(\bm{N})=\mathrm{rank}(\bm{N}^t\bm{N})$, so we can multiply
Eq.~\eqref{finalSystem} by $\bm{N}^t$ preserving the rank and reducing the
number of equations. Reducing the system in the echelon form by applying the
Gauss algorithm~\cite{LEON2009}, we finally obtain 
\begin{equation}
\begin{pmatrix}
\bm{I} & \bm{M} \\
\bm{0} & \bm{0}
\end{pmatrix}
\begin{pmatrix}
\hat{\bm{T}}_{dep}^c \\
\hat{\bm{T}}_{ind}^c
\end{pmatrix}=\bm{0},\qquad \begin{pmatrix}
\hat{\bm{T}}_{dep}^c \\
\hat{\bm{T}}_{ind}^c
\end{pmatrix}=\bm{\Lambda}\hat{\bm{T}}^c,
\end{equation}
where $\bm{\Lambda}$ is the permutation matrix of the Gauss algorithm. In the
previous equation, we split the vector $\hat{\bm{T}}^c$ into two blocks:
$\hat{\bm{T}}_{ind}^c$ is the vector of the independent components while
$\hat{\bm{T}}_{dep}^c= \bm{M}\hat{\bm{T}}_{ind}^c$ is the vector that contains
the dependent entries of $\hat{\bm{T}}$.  The number of dependent entries of
$\hat{\bm{T}}$ is thus given by the rank of $\bm{N}$ and the independent
elements of tensor $\hat{\bm{T}}$ span the kernel of $\bm{N}$.

\section{Explicit form of some relations for the octahedral lattice}\label{sec:explicit_form}

By applying the algorithm presented in Section~\ref{sec:TSymmetriesAlg}, we
derived the explicit form of the local relation Eq.~\eqref{eq:T_definition}
between $\bm{\chi}^{(2)}$ and  $\bm{\zeta}$ that has to be applied in the case
of silicon.  This explicit form is frame dependent and the expressions in
Eq.~\eqref{eq:chi_zeta_crystal} are written with respect to the crystal axis
$x=[100]$, $y=[010]$, $z=[001]$.
\begin{equation}\label{eq:chi_zeta_crystal}
\begin{aligned}
\chi_{111}&=c_1\zeta_{111}+c_4\zeta_{221}+c_4\zeta_{331}+2c_{13}\zeta_{122}+2c_{13}\zeta_{133};\\ 
\chi_{112}&=2c_{11}\zeta_{121}+c_5\zeta_{112}+c_2\zeta_{222} +c_6\zeta_{332}+2c_{12}\zeta_{233};\\ 
\chi_{113}&=2c_{11}\zeta_{131}+2c_{12}\zeta_{232}+c_5\zeta_{113}+c_6\zeta_{223}+c_2\zeta_{333};\\
\chi_{221}&=c_2\zeta_{111}+c_5\zeta_{221}+c_6\zeta_{331}+2c_{11}\zeta_{122}+2c_{12}\zeta_{133};\\
\chi_{222}&=2c_{13}\zeta_{121}+c_4\zeta_{112}+c_1\zeta_{222}+c_4\zeta_{332}+2c_{13}\zeta_{233};\\
\chi_{223}&=2c_{12}\zeta_{131}+2c_{11}\zeta_{232}+c_6\zeta_{113}+c_5\zeta_{223}+c_2\zeta_{333};\\
\chi_{331}&=c_2\zeta_{111}+c_6\zeta_{221}+c_5\zeta_{331}+2c_{12}\zeta_{122}+2c_{11}\zeta_{133};\\
\chi_{332}&=2c_{12}\zeta_{121}+c_6\zeta_{112}+c_2\zeta_{222}+c_5\zeta_{332}+2c_{11}\zeta_{233};\\
\chi_{333}&=2c_{13}\zeta_{131}+2c_{13}\zeta_{232}+c_4\zeta_{113}+c_4\zeta_{223}+c_1\zeta_{333};\\
\chi_{231}&=2c_9\zeta_{231}+2c_{10}\zeta_{132}+2c_{10}\zeta_{123};                              \\
\chi_{232}&=2c_{14}\zeta_{131}+2c_{15}\zeta_{232}+c_7\zeta_{113}+c_8\zeta_{223}+c_3\zeta_{333};\\
\chi_{233}&=2c_{14}\zeta_{121}+c_7\zeta_{112}+c_3\zeta_{222}+c_8\zeta_{332}+2c_{15}\zeta_{233};\\
\chi_{131}&=2c_{15}\zeta_{131}+2c_{14}\zeta_{232}+c_8\zeta_{113}+c_7\zeta_{223}+c_{3}\zeta_{333};\\
\chi_{132}&=2c_{10}\zeta_{231}+2c_9\zeta_{132}+2c_{10}\zeta_{123};                                \\
\chi_{133}&=c_3\zeta_{111}+c_7\zeta_{221}+c_8\zeta_{331}+2c_{14}\zeta_{122}+2c_{15}\zeta_{133};\\
\chi_{121}&=2c_{15}\zeta_{121}+c_8\zeta_{112}+c_3\zeta_{222}+c_7\zeta_{332}+2c_{14}\zeta_{232};\\
\chi_{122}&=c_3\zeta_{111}+c_8\zeta_{221}+c_7\zeta_{331}+2c_{15}\zeta_{122}+2c_14\zeta_{133};\\    
\chi_{123}&=2c_{10}\zeta_{231}+2c_{10}\zeta_{132}+2c_9\zeta_{123}.
\end{aligned}
\end{equation}
The coefficients $c_i$ that appear in this expression are related to the
independent components of $\bm{T}$ by the relations
\begin{equation}
\begin{aligned}
&c_1=\hat{T}_{3333}, \quad c_2=\hat{T}_{2333}, \quad c_3=\hat{T}_{4233}, \quad c_4=\hat{T}_{3323}, \quad c_5=\hat{T}_{2323}, \\
&c_6=\hat{T}_{1323}, \quad c_7=\hat{T}_{5123}, \quad c_8=\hat{T}_{4223}, \quad c_9=\hat{T}_{6363}, \quad c_{10}=\hat{T}_{5263},\\
&c_{11}=\hat{T}_{3153}, \quad c_{12}=\hat{T}_{2153},\quad c_{13}=\hat{T}_{1153}, \quad c_{14}=\hat{T}_{6253}, \quad c_{15}= \hat{T}_{5353}.
\end{aligned}
\end{equation}

The explicit form of the overlap functions to be used in
Eq.~\eqref{eq:effectiveSuscpetibility_ourCase} in the main text are here
reported with respect to the coordinate frames used in
\cite{CHMIELAK2011,CHMIELAK2013} and \cite{DAMAS2014} (the weighted
strain was defined in Eq.~\eqref{eq:weighted_zeta}):

\begin{equation}\label{eq:overlapFactorDef}
\begin{aligned}
o_{1}&=\overline{\zeta^{yy}_{yyy}} \ , 
& o_{2}&=\overline{\zeta^{zz}_{yyy}}+\overline{\zeta^{xx}_{yyy}} \ ,\\
o_{3}&=\R e\left\{\overline{\zeta^{yx}_{xxx}}+\overline{\zeta^{yx}_{zzx}}\right\} \ ,
& o_{4}&=\overline{\zeta^{yy}_{xxy}}+\overline{\zeta^{yy}_{zzy}} \ , \\
o_{5}&=\frac{1}{2}\overline{\zeta^{xx}_{xxy}} +\frac{1}{2}\overline{\zeta^{zz}_{zzy}} +\frac{1}{2}\overline{\zeta^{zz}_{xxy}} +\frac{1}{2}\overline{\zeta^{xx}_{zzy}} \ , 
& o_6&=o_5 ,\\
o_{7}&=o_{3}, 
& o_{8}&=2\R e\left\{\overline{\zeta^{xy}_{yyx}}\right\}\ , \\
o_{9}&=\overline{\zeta^{xx}_{xxy}}+\overline{\zeta^{zz}_{zzy}}-\overline{\zeta^{zz}_{xxy}}-\overline{\zeta^{xx}_{zzy}} \ , 
& o_{10}&=2\overline{\zeta^{xx}_{xyx}}-2\overline{\zeta^{zz}_{xyx}}\ , \\
o_{11}&=\overline{\zeta^{xx}_{xyx}}-\overline{\zeta^{zz}_{xyx}}\ ,  
& o_{12}&=o_{11} \ , \\
o_{13}&=2\overline{\zeta^{yy}_{xyx}}\ , 
& o_{14}&=2\R e\left\{\overline{\zeta^{yx}_{xxx}} -\overline{\zeta^{yx}_{zzx}}\right\}\ , \\
o_{15}&=4\R e\left\{\overline{\zeta^{yx}_{xyy}}\right\}.
\end{aligned}
\end{equation}

\section{The variation of the refraction index}\label{sec:neff_change}

In this appendix we report some details on the expression for the variation of
the effective refraction index induced by a non-vanishing $\bm{\chi}^{(2)}$,
following an argument analogous to the one used in
Refs.~\cite{CHEN2006}. 

Let us consider the following form of the reciprocity theorem for a waveguide
(see e.g. \cite{SNYDER1983} \S 31-1):
\begin{equation}\label{eq:rth_conj}
\frac{\partial}{\partial z} \int_{A_{\infty}}\bm{F}_c\cdot\bm{i}_z\mathrm{d} A=
\int_{A_{\infty}}\bm{\nabla}\cdot \bm{F}_c\,\mathrm{d} A\ ,
\end{equation}
where the waveguide is directed along the $z$ direction, $A_{\infty}$ is the
plane orthogonal to it and the vector $\bm{F}_c$ is defined by 
\begin{equation}\label{F_def}
\bm{F}_c=\bm{E}_0^*\times\bm{H}+\bm{E}\times\bm{H}_0^*\ .
\end{equation}
In the following $\bm{E}_0$ and $\bm{H}_0$ will be the optical fields
propagating in the guide when $\bm{\chi}^{(2)}\equiv 0$, while $\bm{E}$ and
$\bm{H}$ will be the corresponding fields when the nonlinear susceptibility is
non-vanishing.  It is simple to show that, when all the field have angular
frequency $\omega$, Eq.~\eqref{eq:rth_conj} can be rewritten in the form
\begin{equation}\label{eq:rth_final}
\frac{\partial}{\partial z}\int_{A_{\infty}}(\bm{E}_0^*\times\bm{H}+\bm{E}\times\bm{H}_0^*)
\cdot\bm{i}_z\mathrm{d} A = i\omega\int_{A}\bm{E}_0^*\cdot\bm{P}^{(2)}\mathrm{d} A\ ,
\end{equation}
where $\bm{P}^{(2)}$ is the polarization induced by the nonlinear
susceptibility ($\bm{P}=\bm{P}_0+\bm{P}^{(2)}$) and $A$ is the section of the
wave guide in which $\bm{\chi}^{(2)}$ is non-vanishing. 

We will use for the unperturbed fields the form
\begin{equation}
\begin{aligned}
& \bm{E}_0(\bm{r}, t)=\bm{e}_0(x,y; \omega_0)e^{i(k_0 z-\omega t)} \ , \\
& \bm{H}_0(\bm{r}, t)=\bm{h}_0(x,y; \omega_0)e^{i(k_0 z-\omega t)} \ ,
\end{aligned}
\end{equation}
while we will assume for the perturbed fields the expressions
\begin{equation}\label{eq:fields}
\begin{aligned}
& \bm{E}(\bm{r}, t)=u(z)\bm{e}_0(x,y; \omega_0)e^{i(k z-\omega t)}\ , \\
& \bm{H}(\bm{r}, t)=u(z)\bm{h}_0(x,y; \omega_0)e^{i(k z-\omega t)}\ ,
\end{aligned}
\end{equation}
thus assuming that the nonlinearity does not significantly affect the transverse modes.

Using these forms of the fields in Eq.~\eqref{eq:rth_final}, together with
Eq.~\eqref{eq:pockels_p}, which can be written as
\begin{equation}
\bm{P}^{(2)}=2\epsilon_0\bm{\chi}^{(2)}(\omega;\omega,0):\bm{E}\bm{E}^{dc}\ , 
\end{equation}
we obtain a differential equation for the envelope function $u(z)$, namely:
\begin{equation}
\frac{\partial u(z)}{\partial z}+i(k-k_0)u(z)=i\omega u(z) X\ ,
\end{equation}
where we introduced the notation
\begin{equation}
X=
\frac{2\epsilon_0\int_{A}\bm{e}_0^{\,*}\cdot\bm{\chi}^{(2)}:\bm{e}_0\bm{E}^{dc}\,\mathrm{d} A}
{\int_{A_{\infty}}(\bm{e}_0\times\bm{h}_0^*+\bm{e}_0^{\,*}\times\bm{h}_0)\cdot
\bm{i}_z\,\mathrm{d}A}\ .
\end{equation}
By solving this equation and inserting the solution in Eq.~\eqref{eq:fields} we
finally obtain $k=k_0+\omega X$, thus corresponding to the variation $\Delta
n^{\mathrm{eff}}=c X$ of the effective refraction index.

\section*{Acknowledgments}
This work is partially supported by the Italian Ministry of Education,
University and Research (MIUR) through the FIRB project ``MINOS''. The authors
would like to thank Martino Bernard, Massimo Borghi, Mattia Mancinelli, and
Lorenzo Pavesi from University of Trento; Mher \replaced{Ghulinyan}{Gullynham}, Georg Pucker from
Fondazione Bruno Kessler; Nicola Andriolli, Isabella Cerutti,  Koteswararao Kondepu and Andrea Merlo from 
Scuola Superiore Sant'Anna;  Giuseppe Rodriguez from University of Cagliari for
the useful discussions and support. It is a pleasure to thank especially
Fabrizio Di Pasquale from Scuola Superiore Sant'Anna for his beautiful
collaboration and insightful advice.

\newpage
%\bibliographystyle{apsrev4-1}
%\bibliography{bibliography}

% that's all folks
\end{document}